\RequirePackage{fix-cm}
\documentclass[smallextended]{svjour3}       
\smartqed  
\usepackage{graphicx}
\usepackage{amsmath}
\usepackage{bbm}
\usepackage{algorithm}
\usepackage{algorithmicx}
\usepackage{algpseudocode}
\usepackage{arydshln}
\usepackage[normalem]{ulem}

\usepackage[numbers,square,sort]{natbib}
\usepackage[caption=false]{subfig}
\usepackage{csvsimple}
\usepackage{tabularx}
\usepackage{booktabs,caption}
\usepackage[flushleft]{threeparttable}
\usepackage{xcolor,colortbl}

\def\tsc#1{\csdef{#1}{\textsc{\lowercase{#1}}\xspace}}
\tsc{WGM}
\tsc{QE}

\newcommand{\vx}{{\mathbf x}}

\newif\ifnotesw \noteswtrue

\noteswfalse   

\newcommand{\orl}{$\mathit{orl}$ }
\newcommand{\pp}{$p^\prime$}

\journalname{Computational Geosciences}
\begin{document}

\title{Impact of artificial topological changes on flow and transport through fractured media due to mesh resolution}



\titlerunning{Impact of false connections in ECPMs}        

\author{Aleksandra A. Pachalieva \and
        Matthew R. Sweeney	\and
	    Hari Viswanathan \and
	    Emily Stein \and
	    Rosie Leone \and
	    Jeffrey D. Hyman
}


\institute{ Aleksandra A. Pachalieva \at
            Center for Nonlinear Studies \\
            Theoretical Division \\
            and \\
            Computational Earth Science Group (EES-16) \\
            Earth and Environmental Sciences Division \\
            Los Alamos National Laboratory \\
            Los Alamos, NM 87544, USA \\
            \email{apachalieva@lanl.gov}           
            \and
	    Matthew R. Sweeney \at
            Computational Earth Science Group (EES-16) \\
            Earth and Environmental Sciences Division \\
            Los Alamos National Laboratory \\
            Los Alamos, NM 87544, USA \\
            \email{msweeney2796@lanl.gov}           
            \and
            Hari Viswanathan \at
            Computational Earth Science Group (EES-16) \\
            Earth and Environmental Sciences Division \\
            Los Alamos National Laboratory \\
            Los Alamos, NM 87544, USA \\
            \email{viswana@lanl.gov}
            \and
            Emily Stein \at
            Nuclear Fuel Cycle Technologies \\
            Sandia National Laboratories \\
            Albuquerque, NM 87123, USA \\
            \email{ergiamb@sandia.gov}
            \and
            Rosie Leone \at
            Nuclear Fuel Cycle Technologies \\
            Sandia National Laboratories \\
            Albuquerque, NM 87123, USA \\
            \email{rleone@sandia.gov}
            \and
            Jeffrey D. Hyman \at
            Computational Earth Science Group (EES-16) \\
            Earth and Environmental Sciences Division \\
            Los Alamos National Laboratory \\
            Los Alamos, NM 87544, USA \\
            \email{jhyman@lanl.gov}
}

\date{Received: date / Accepted: date}

\maketitle


\begin{abstract}
We performed a set of numerical simulations to characterize the interplay of fracture network topology, upscaling, and mesh refinement on flow and transport properties in fractured porous media.
We generated a set of generic three-dimensional discrete fracture networks at various densities, where the radii of the fractures were sampled from a truncated power-law distribution, and whose parameters were loosely based on field site characterizations. 
We also considered five network densities, which were defined using a dimensionless version of density based on percolation theory. 
Once the networks were generated, we upscaled them into a single continuum model using the upscaled discrete fracture matrix model presented by~\citet{sweeney2020upscaled}.
We considered steady, isothermal pressure-driven flow through each domain and then simulated conservative, decaying, and adsorbing tracers using a pulse injection into the domain. 
For each simulation, we calculated the effective permeability and solute breakthrough curves as quantities of interest to compare between network realizations.
We found that selecting a mesh resolution such that the global topology of the upscaled mesh matches the fracture network is essential. 
If the upscaled mesh has a connected pathway of fracture (higher permeability) cells but the fracture network does not, then the estimates for effective permeability and solute breakthrough will be incorrect. 
False connections cannot be eliminated entirely, but they can be managed by choosing appropriate mesh resolution and refinement for a given network. 
Adopting octree meshing to obtain sufficient levels of refinement leads to fewer computational cells (up to a 90\% reduction in overall cell count) when compared to using a uniform resolution grid and can result in a more accurate continuum representation of the true fracture network.

\keywords{Discrete fracture network \and Continuum model \and Subsurface flow and transport \and Upscaling \and Computational geometry \and Mesh generation}
\end{abstract}

\section{Introduction}
\label{sec:introduction}

Characterizing the flow of fluids and the associated transport of dissolved chemicals through subsurface fractured porous media is a critical step for many industrial and civil engineering endeavors, including the geological sequestration CO$_2$~\cite{hyman2019characterizing,jenkins2015state}, drinking water aquifer management~\citep{kueper1991behavior,viswanathan2022from}, the environmental restoration of contaminated fractured media~\citep{national1996rock,neuman2005trends,vanderkwaak1996dissolution}, hydrocarbon extraction~\citep{hyman2016understanding,middleton2015shale}, and the permanent disposal of spent nuclear fuel and high-level radioactive waste~\citep{hadgu2017comparative,joyce2014multiscale}.
A cornerstone of simulating flow, transport, or heat transfer in fractured porous media is the adopted system representation, a computational mesh and solving the governing equations of interest on said mesh.
Recent advances in discrete representations of fractures, such as discrete fracture network (DFN) and discrete fracture matrix (DFM) models, have vastly improved our capabilities to model physical and chemical processes on realistic fracture networks, but are still limited in several regards.
DFN models only represent the fractures, neglecting important contributions from the surrounding rock (i.e., matrix), which can be critical in a variety of applications, including enhanced geothermal and unconventional reservoirs. DFM models are unequivocally the most accurate representation of the true fractured system~\citep{berre2019flow}, but are still limited by difficulties in meshing algorithms and solving the governing equations, which is complicated by the coupling between fracture and matrix domains. 
While DFN models have been used to model tens of thousands of fractures, DFM models have only been used to model hundreds~\cite{hyman2022flow}.
Consequently, we frequently employ continuum models, also known as equivalent continuum/continuous porous medium models (ECPM), where the fracture network properties are upscaled into equivalent properties of a porous medium when the fracture number is large and the matrix processes cannot be neglected. 
They are typically used due to their ease of implementation and wide compatibility with existing multi-physics codes. However, they are known to smear out important length scales and it is not fully understood whether or how their accuracy depends on the underlying fracture network properties and the resolution of the computational mesh in relation. 
Previous work by~\citet{jackson2000} showed that ECPMs are able to represent the underlying DFNs, in terms of reproducing the effective permeability.
However, they only considered two different well-connected DFNs of relatively high density and no flow contribution from the matrix. 
More recently,~\citet{kottwitz} systematically analyzed the relationship between the resolution of the ECPM mesh and the effective permeability in steady state flow simulations. They found monotonic convergence of the effective permeability as the mesh size was decreased and suggest minimum and maximum cell sizes related to fracture sizes needed in continuum meshes to preserve the underlying network characteristics. The networks considered in their study were networks with higher fracture intensity/density.
\citet{sweeney2020upscaled} also showed that upscaling for a single fracture is accurate regardless of mesh size. 
Taken together with prior studies, this suggests that changing the topology of the effective network is responsible for differences when considering different mesh resolutions for the same fracture network. For instance, at a coarse resolution, properties of two fractures may be upscaled to the same cell in the continuum mesh, even if they do not intersect in the underlying network, in effect creating a "false connection" in the network.
These types of changes in topology have never been systematically investigated, but are critical to understanding whether we can actually use continuum models with confidence for a spectrum of DFNs, not just ones with high intensity/density.
In principle, if two fractures are close to one another and do not intersect, then the meshing generation parameters need to be selected so that the background element sizes are smaller than the distance between the fractures to eliminate a false connection.  
Within a stochastically generated fracture network, attempts to resolve these distances, which will quickly become below the size resolvable with finite precision arithmetic, are futile regardless of the heroic efforts put forth. 
Thus, false connections are unavoidable when upscaling a stochastically generated network.
Moreover, multiple false connections can occur within a single control volume.  
Deviation of connectivity/topology of the continuum mesh from the fracture network could influence the local and global flow and transport within the network.
Specifically, global changes in topology could occur when false connections result in a continuum mesh with a connected pathway of fracture cells between flow boundaries, i.e., percolates, when the underlying DFN does not percolate. 
However, characterizing the impact of these false connections has yet to be systematically performed, and and general guidelines have yet to be established. 

In this work, we have assessed the impacts of so-called false connections on the accuracy and convergence of flow and transport in fractured systems represented by an ECPM model. 
We found that selecting a mesh resolution such that the global topology of the upscaled mesh matches the fracture network is essential. 
If the upscaled mesh has a connected pathway of fracture (higher permeability) cells but the fracture network does not, then the estimates for effective permeability and solute breakthrough will be incorrect. 
Local false connections between fractures due to a coarse mesh result in more solute dispersion, but to a smaller degree than if there is mismatch in global connectivity, which is consistent with prior work.
To this end, we provide some principle-based guidelines.
Foremost, macro-scale connectivity is an essential property of the network to be represented in an upscaled continuum.   
Likewise, while false connections cannot be eliminated entirely, they can be managed by choosing appropriate mesh resolution and refinement for a given network to obtain more accuracte estimations of effective permeability and breakthrough curve behavior.
These general guidelines are refined both in terms of the difference between matrix and fracture permeabilities and network densities.


\section{Numerical Simulations}
Fracture network generation, upscaling, flow and transport simulations are performed using the {\sc dfnWorks} computational suite~\citep{hyman2015dfnworks}. 
{\sc dfnWorks} combines the feature rejection algorithm for meshing {\sc fram}~\citep{hyman2014conforming,krotz2021maximal}, the LaGriT meshing toolbox~\citep{lagrit2013}, and the parallelized subsurface flow and reactive transport code {\sc pflotran}~\citep{pflotran-user-ref}. 
{\sc fram} is used to generate three-dimensional fracture networks. 
LaGriT is used to create a computational mesh representation of the DFN in parallel. 
The networks are upscaled to an equivalent continuum porous media using the upscaled discrete fracture matrix (UDFM) model of Sweeney et al.~\citep{sweeney2020upscaled}. 
{\sc pflotran} is used to numerically integrate the governing equations to obtain steady state flow field fields within the domains, and then simulate transport therein. 

\subsection{Network Generation}

We generate a set of generic three-dimensional fracture networks at various densities. 
The networks are generic in that they do not represent a particular field site, but their parameters are based on field site characteristics of fractured crystalline rock~\citep{bonnet2001scaling}. 
The networks are generated in a cubic domain with sides of length  $L = 50$ meters. 
All networks are generated using the same parameters, except the number of fractures, which we use to control the network density. 
The networks consist of a single fracture family. 
Each fracture is a planar disc with an aspect ratio of one, i.e., each fracture is a circle. 
The radius $r$ of the fractures are sampled from a truncated power-law distribution with a probability density function of
\begin{equation}\label{eq:tpl}
p_r(r,r_0,r_u) = \frac{\alpha}{r_0} \frac{(r/r_0)^{-1-\alpha}}{1 - (r_u/r_0)^{-\alpha}},
\end{equation}
and parameters: decay exponent $\alpha = 1.8$, lower cut off $r_0= 1$, and upper cut off $r_u = 10$m.
This style DFN generation is referred to as a Poissonian generation opposed to a kinematic generation where fractures nucleate and grow~\cite{davy2010likely,davy2013model,lavoine2020discrete,thomas2020permeability}.
Fracture centers are uniformly distributed throughout the domain. 
During the generation of the network, we expand the domain slightly (5 meters) in each direction to avoid low-density issues near domain boundaries.

The fracture orientations are uniformly distributed across the unit sphere,  which mimics disordered fractures networks observed in the field~\citep{hyman2018dispersion,klint2004multi}. 
The orientations are sampled from a three dimensional von Mises Fisher distribution,
\begin{equation}\label{eq:fisher}
f({\bf x}; {\boldsymbol \mu}, \kappa ) = \frac{ \kappa \exp( \kappa {\boldsymbol \mu}^{T} {\bf x} )}{ 4 \pi \sinh(\kappa)}~.
\end{equation}
In \eqref{eq:fisher},  ${\boldsymbol \mu}$ is the mean direction vector of the fracture family, $T$ denotes transpose, and $\kappa \geq 0$ is the concentration parameter that determines the degree of clustering around the mean direction. 
Values of $\kappa$ close to zero lead to a uniform distribution of points on the sphere while larger values create points with a small deviation from mean direction.
To obtain uniformly random orientations, we set $\kappa = 0.1$ and mean normal vector of $(0,0,1)$.
The distribution is sampled using the method detailed in~\cite{wood1994simulation}.

Variations in the hydraulic properties of the fractures are created by positively correlating each fracture's hydraulic aperture to its radius via
\begin{equation}\label{eq:aperture}
    b = 5.0 \times 10^{-4} \sqrt{r}\,.
\end{equation}
The parameters of the relationship are based on field data~\citep{SKB2010}, again in a fractured crystalline rock.
This perfectly correlated relationship between fracture size and hydraulic aperture is a common choice for DFN models~\citep{bogdanov2007effective,dreuzy2002hydraulic,frampton2010inference,hyman2016fracture, joyce2014multiscale,wellman2009effects}. 
Under this model setup, the hydraulic aperture is constant within each fracture, but it varies between fractures with larger fractures having wider hydraulic apertures than smaller fractures. 
We do not include hydraulic aperture variability into these simulations because our goal is to upscale the networks to an equivalent continuum, and such properties are integrated out. 

\begin{figure}[t]
     \centering
     \includegraphics[width=\columnwidth]{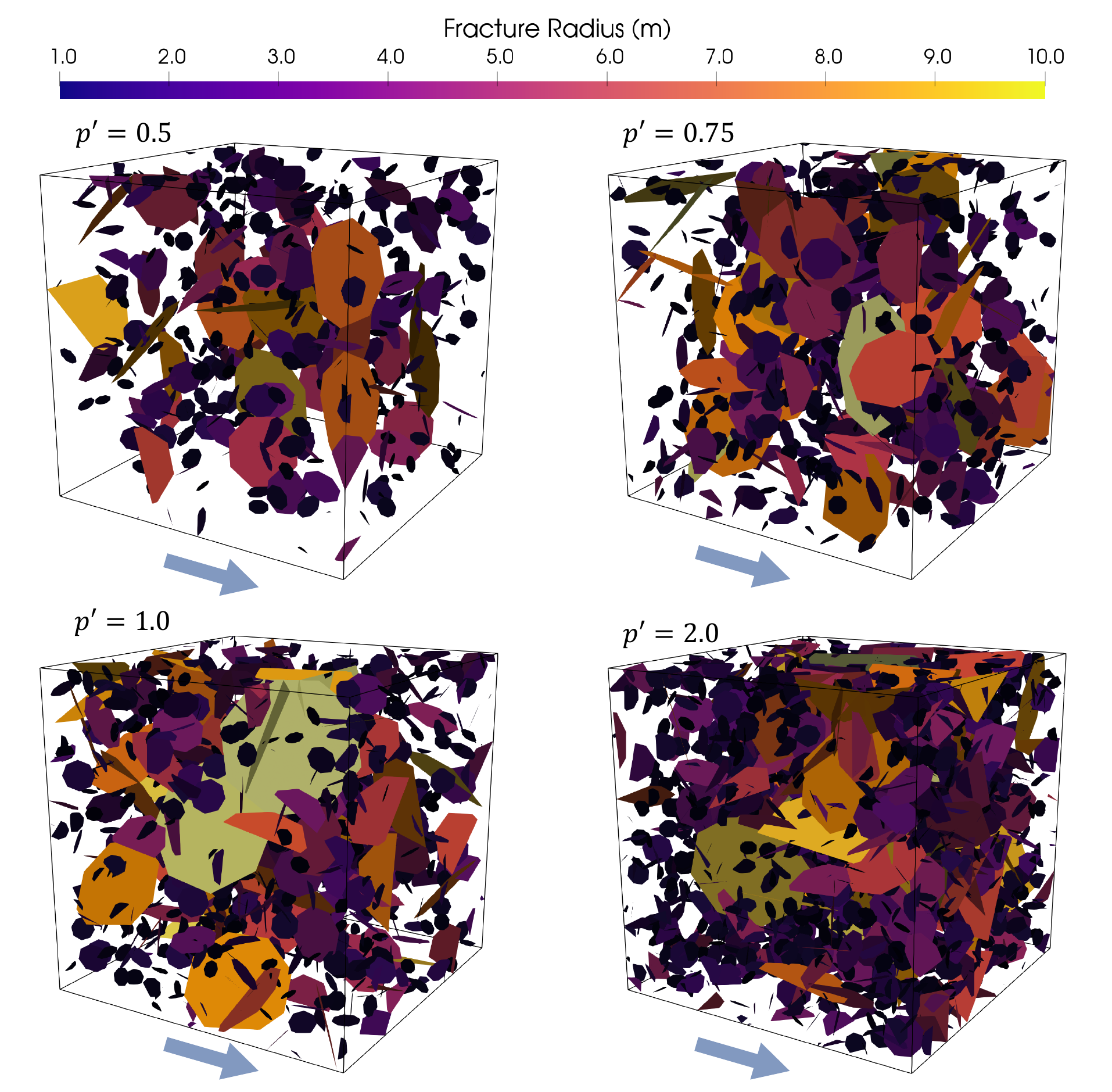}		   
     \caption{Network realizations from each density considered in this publication ($p^\prime = 0.5, 0.75, 1.0$ and $2.0$). The fracture radius is color-coded to highlight the range of length scales in the simulations. When the network's density is below the percolation threshold ($p^\prime < 1$), the fracture network often does not connect between the inflow and outflow boundaries. For networks above the percolation threshold ($p^\prime > 1$), we can clearly see at least one path connecting the inflow and outflow boundaries. As the density increases, the number of fractures increases and thus the number of possible paths though the fracture network.}
     \label{fig:fracture_radius}
\end{figure}

The set of DFNs is ordered by network density. 
We adopt a definition of network density based on the percolation parameter $p$ provided by~\citet{dreuzy2012influence}, which is defined as
\begin{equation}\label{eq:density}
p = \frac{N}{L^2} \int_{r_0}^{r_u} dr \min(r,\alpha L) p_r(r)~.
\end{equation}
We define the critical percolation density value (denoted $p_c$) to be the minimum number of fractures required so there is almost surely a connected cluster of fractures that spans the whole domain, i.e., the value of $N$ such that $p$ in \eqref{eq:density} is equal to unity~\citep{berkowitz1993percolation,bour1997connectivity,bour1998connectivity,sahimi1994applications}.
For the domain size and truncated power-law distribution parameters considered here, $p_c \approx 1000$ fractures. 
This definition of density provides a normalization factor so that networks can be placed in a relative context using a dimensionless percolation parameter
\begin{equation}\label{eq:p_prime}
p^\prime = p/p_c\,,
\end{equation}
where $p$ is the number of fractures in a particular network during generation. 
We consider values of $p^\prime = 0.5, 0.75, 1.0, 1.5$ and $2.0$ (Figure~\ref{fig:fracture_radius}).
The fractures are colored by their radius to demonstrate the wide range of length scales in the simulations.

We consider two sub-cases for each density value.
In the first, we consider the network with all of the generated fractures.
In the second, we only consider isolated fractures and clusters relative to the flow conditions. i.e., we retain only clusters of fractures that connect the inflow and outflow boundaries.
We refer to these cases as isolated fractures retained and isolate fractures removed, respectively.
The second case, where isolated fractures are removed, is a common pre-processing step in DFN modeling used to reduce the computational cost of the simulation. 
These isolated sub-networks do not contribute to network flow, i.e., simulating flow within only the fracture network domain, and can therefore be removed without influencing the simulation. 
However, the impact of removing/including isolated fractures within the upscaled system is relatively un-characterized, and performing this characterization is a goal of this study.
Detecting clusters that connect the domain and isolated clusters is performed using a graph-based algorithm within dfnWorks, cf.~\cite{hyman2017predictions} and~\cite{hyman2018identifying} for details of the graph-based techniques.
If a network connects the inflow and outflow boundaries, then we say the network is {\it percolating}, else the network is {\it non-percolating}.
The $p^\prime = 0.5$ and $0.75$ networks are non-percolating, while all other networks $p^\prime \geq 1$ are percolating.

Table \ref{table:network_params} reports the dimensionless network density ($p^\prime$), initial number of fractures ($N$), number of non-isolated fractures ($\widehat{N}$), percentage of the total number of fractures connecting the inflow/outflow boundaries ($\widehat{N}/N$), fracture intensity with isolated fractures retained ($P_{32}$),  and fracture intensity with isolated fractures removed ($\widehat{P}_{32}$).
The fracture intensity is defined as 
\begin{equation}
    P_{32} = \frac{1}{V}\sum S_f,
\end{equation}
where $S_f$ is fracture surface area and the sum is taken over all fractures in the network, and V is the total volume of the domain.
$P_{32}$ is reported in units of m$^{-1}$. 
Since all network generation parameters are fixed, the initial number of fractures placed in the domain increases in proportion to the selected density, cf.~\eqref{eq:p_prime}.
In the case of $p^\prime = 0.5$ and $0.75$, there are no fractures remaining in the domain once isolated fractures have been removed and $P_{32} = 0$.
Therefore, we only consider the cases where all fractures are retained for those two densities, as the upscaled continuum media would otherwise be uniformly matrix cells.  
For the percolating networks, $p^\prime \geq 1$, the percentage of the initial network retained increases slightly with density, but not in the same linear fashion as the total number of fractures initially placed into the domain. 
Likewise, the fracture intensity increases once the network percolates.  

\begin{table}[b]
\centering
\caption{Network Characterization: dimensionless network density ($p^\prime$), initial number of fractures ($N$), number of non-isolated fractures ($\widehat{N}$), percentage of the total number of fractures connecting the inflow/outflow boundaries ($\widehat{N}/N$), fracture intensity with isolated fractures retained ($P_{32}$), fracture intensity with isolated fractures removed ($\widehat{P}_{32}$).}\label{table:network_params}
\begin{tabular}{|p{0.06\linewidth}p{0.12\linewidth}p{0.12\linewidth}p{0.12\linewidth}p{0.12\linewidth}p{0.12\linewidth}|}
\hline
$p^\prime$ & $N$ & $\widehat{N}$ & $\widehat{N}/N$  & $P_{32}$ [m$^{-1}$] & $\widehat{P}_{32}$ [m$^{-1}$] \\ \hline
0.50 & 500 & 0    & 0.0\,$\%$ & 0.128 & 0  \\  
0.75 & 750  & 0    & 0.0\,$\%$  &  0.211   &   0 \\  
1.00 & 1000 & 68   & 6.8\,$\%$ & 0.261 & 0.046 \\
1.50 & 1500 & 213  & 14.2\,$\%$  & 0.425 & 0.124 \\ 
2.00 & 2000 & 286  & 14.3\,$\%$ & 0.492 &  0.131  \\\hline
\end{tabular}
\end{table}

\subsection{UDFM}
Once the networks are generated, we upscale them into a single continuum model using the UDFM method presented by~\citet{sweeney2020upscaled}.
There are two main steps to the UDFM method, which are (1) creating the variable resolution continuum mesh and (2) combining the hydraulic attributes of the fractures, e.g., apertures; surface areas; and volumes, with the background matrix permeability and porosity values.
We highlight the main aspects of each step below.
Complete details of the model along with verification examples are provided in~\cite{sweeney2020upscaled}.

\subsubsection{Meshing}
Given a DFN, the UDFM method uses octree refinement to produce a spatially variable Delaunay tetrahedral mesh where the mesh resolution is a function of the distance to the fractures in the DFN, i.e., the mesh is refined more in closer proximity to fractures.  
The higher resolution close to the fractures allows for the representation of small length scales, which, in turn,  allows for the resolution of local gradients in flow and transport simulations, the highest of which occur close to the fractures. 
The control parameters for the mesh generation are the edge length of the initial mesh element size ($l$) and the number of refinement levels desired in the final continuum mesh (\orl).
The initial mesh size is also the largest in the domain.
The meshing algorithm initializes a uniform hexahedral mesh with edge length $l$. 
The cells in the initial mesh that intersect fractures in the DFN are tagged as {\it fracture} cells, while the cells that are not intersected by a fracture are tagged as {\it matrix} cells. 
All of the fracture cells and their neighbors (cells that share a face with a fracture cell) are refined into eight sub-hexagons with edge length $l/2$ to create a child mesh.
This process is performed recursively until the desired \orl is obtained.
The final refined hexahedral mesh is then converted to a Delaunay tetrahedral mesh, whose dual mesh is a Voronoi tesselation that can be used in two-point flux simulation codes.

\begin{figure}
     \centering
     \includegraphics[width=1.0\columnwidth]{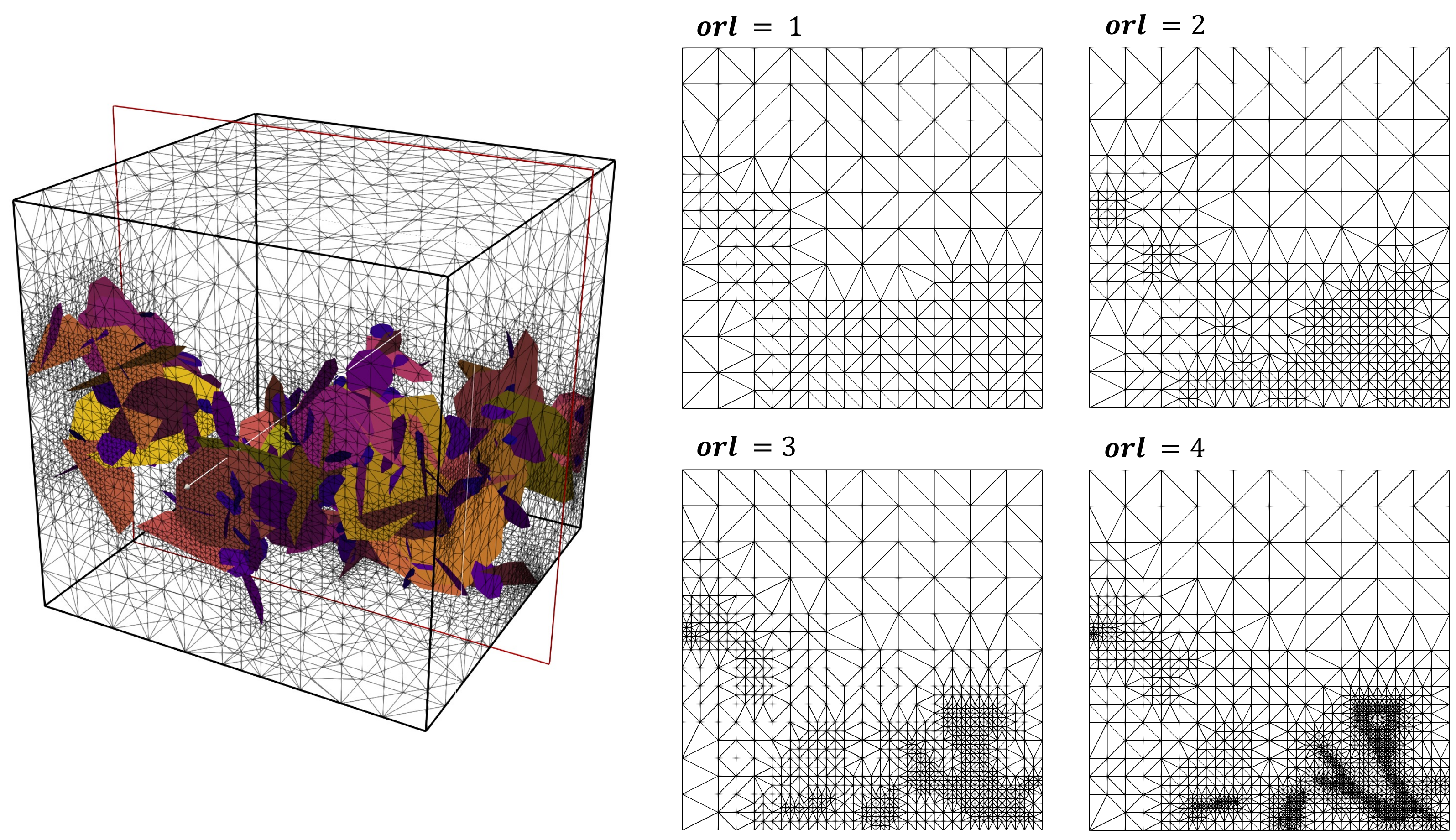}
     \caption{Mesh refinement levels applied to a fracture network with density \pp$ = 1.5$. As the octree refinement level increases, we can see the individual fractures getting better resolved. This makes the meshing more computationally demanding, but reduces the number of false connections and allows for more accurate representation of the flow and transport properties of the system. }
     \label{fig:diff_orls}
\end{figure}

For these simulations we fix $l = 5$ m  and solely consider the impact of varying \orl values.
It should be noted that different choices of $l$ and \orl can give the same resolution near the fractures. For example, a mesh generated with values of $l = x$ and \orl$ = y$ will have the same finest mesh resolution as a mesh generated with values of $l=2x$ and \orl$ = y+1$, or, $l = x/2$ and \orl$ = y-1$, but not necessarily the same total number of cells. Generally, it is best to balance the computational load of meshing with that of simulation, which means the user must determine {\it a priori} the appropriate contribution of each parameter to the resolution and refinement, which itself might change depending on the network. From prior work, we know going beyond \orl$ = 4$ creates a bottleneck in the meshing algorithm, but starting with an $l$ too small will defeat the purpose of the refinement to begin with, creating a mesh with too many cells for typical simulation tools~\citep{sweeney2020upscaled}.
Consequently, we select octree refinement levels of \orl $ = 1, 2, 3,$ and $4$.
Figure~\ref{fig:diff_orls} shows the mesh for the \pp$ = 1.5$ network.
As the octree refinement level increases, the fractures are better resolved in the continuum mesh. 
While larger values of \orl are more computationally demanding, they allow for a better continuum representation of the DFN.

\subsubsection{False Connections}

\begin{figure}
     \centering
     \includegraphics[width=0.7\columnwidth]{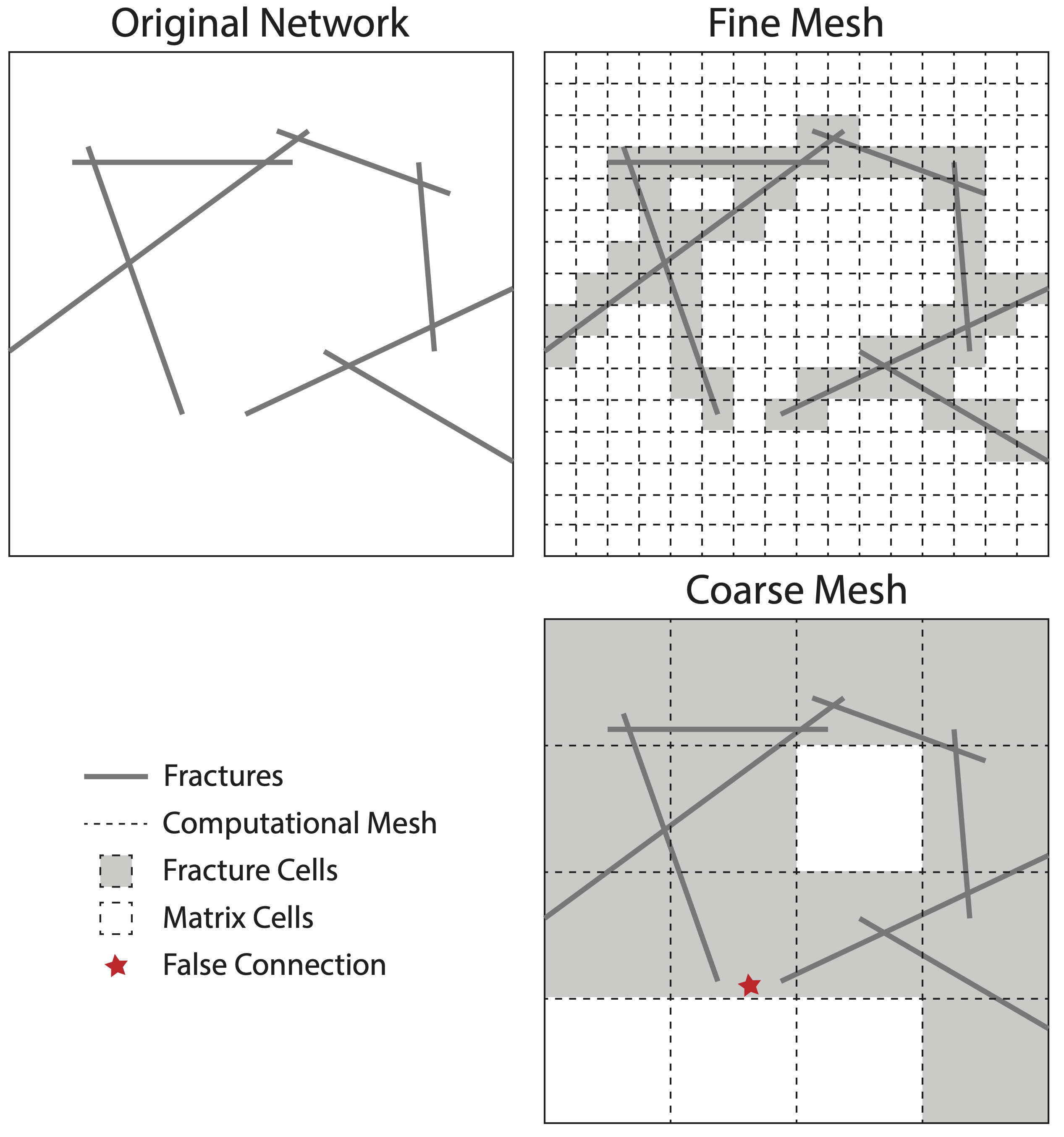}
    \caption{Example of how false connections can change connectivity in ECPM mesh. In the original network, there is only one percolating pathway connecting the left and right boundaries. When a fine ECPM mesh is used to represent the network, the correct connectivity is preserved, whereas when a coarser ECPM mesh is used to represent the network, an artificial connection appears in the lower half of the domain.}
    \label{fig:false_connections}
\end{figure}

When mapping a DFN to an ECPM, it is possible to create \emph{false connections} between fractures. 
We define a false connection in the continuum mesh as when two fractures that do not intersect in the DFN are upscaled into the same control volume cell in the continuum mesh. 
Figure~\ref{fig:false_connections} presents a two-dimensional pictorial explanation of a false connection.
The original network of fractures, grey lines, is shown in the upper left sub-figure.
There is only one pathway in the network between the left and right domain boundaries. 
The upper right sub-figure shows a continuum mesh where fracture cells in the mesh are grey and matrix cells are white. 
Note that in this cartoon, a uniform quad mesh is used for simplicity. 
In this case, the underlying topology of the DFN is preserved by the upscaled mesh. 
The lower right sub-figure shows another continuum mesh with coarser resolution.
Here, a false connection is created in the bottom of the domain, indicated by a red star, which allows for fracture flow along two pathways rather than one. 

A primary design goal of the UDFM method is to minimize the occurrence of false connections, through the use of spatially variable mesh resolution, while also minimizing the number of mesh elements, and in turn computational cost of the simulation.  
These two desires compete with one another. 
Selecting a value of \orl allows for some control of the number of false connections, but cannot remove them entirely.
The advantage of the UDFM method compared to a uniform resolution mesh is that to achieve the same number of false connections in the domain requires drastically fewer elements, due to the variable refinement. 
Table~\ref{table:octree_fc} reports false connections and mesh information for the different refinement levels across all density values. 
The total number of false connection (\#f), the number of cells with false connections (fc), the total number of Voronoi cells (vc), the percentage of Voronoi cells with false connections (fc/vc $\times$ 100). 
For comparison, the number of cells in an equivalent uniform resolution hexahedral mesh (n) are also included along with a computational cost comparison (vc/n $\times$ 100) in terms of reduction of number of mesh cells, which in turn reduces the degrees of freedom in the computational physics simulation and over all cost.
By equivalent uniform resolution hexahedral we mean meshing the domain with  hexahedrons of size equal to the smallest hexahedral produced using the octree refinement level. 
The equivalent number of hexahedral cells ($n$) is given by
\begin{equation}
n = \left (\frac{L}{\delta x}+ 1 \right )^3, \qquad \text{where} \qquad \delta x = \frac{l}{2^{\mathit{orl}}}\, ,
\end{equation}
where $L$ is the domain size length, L = 50. 
The plus one in the first expression is for end points of the domain edges. 

As \orl increases, the number of false connections decreases monotonically, as expected.
The difference in the number of false connections between subsequent \orl values decreases with increasing \orl values, i.e., the difference between \orl $= i$ and \orl $= i+1$ is larger than the difference between \orl $= i+1$ and \orl $= i+2$.
This is true for all observed densities, when isolated fractures are both retained and removed.
Removal of isolated fractures also impacts the number of false connections. 
In the case of retaining the isolated fractures, there are higher percentages of Voronoi cells that contain a false connection when compared to the cases where isolated fractures are removed. 
However, the difference between the two cases reduces as \orl increases.
By \orl $= 3$, the two values are quite similar.
Likewise, the UDFM mesh size relative to the equivalent hexahedral mesh is comparable when isolated fractures are retained at low \orl values, but significantly smaller when isolated fractures are removed.
Even for \orl $= 1$, the UDFM mesh has 50\% fewer cells than the equivalent hexahedral mesh. 
As \orl increases the UDFM meshes reduce in size relative to the equivalent hexahedral mesh even more, for both cases.
Network density also influences the number of false connections along with the \orl values.
As the density increases, the number of false connections increases, which is expected as there are more fractures in the domain, and they can be closer to one another.

\begin{table}[h]
\centering
\caption{UDFM mesh information for network densities \pp at different \orl values where $n$ is the equivalent number of hexahedral cells.
Number of false connections (\#f), number of cells with false connections (fc), total number of Voronoi cells (v), percentage of Voronoi cells with false connections (fc/vc[\%]), and relative mesh size compared to equivalent hexahedral mesh (vc/n[\%]).}
\label{table:octree_fc}
\begin{tabular}{l|rrrrr|rrrrr}
 \multicolumn{11}{c}{\orl = 1, $n = 9,261$} \\ \hline
 &\multicolumn{5}{c}{Isolated Fractures Retained} &\multicolumn{5}{c}{Isolated Fractures Removed} \\
\hline
$p^\prime$ & \#f & fc & vc & fc/vc[\%] & vc/n[\%] & \#f & fc & vc & fc/vc[\%]  & vc/n[\%]  \\
\hline
0.50 & 1277 & 2414 & 9106 & 26.51 & 98.33         & - & - & - & - & - \\
0.75 & 2892 & 4193 & 9254 & 45.31 & 99.92        & - & - & - & - & - \\
1.00 & 5051 & 5472 & 9261 & 59.09 & 100.00         & 178 & 546 & 2894 & 18.87 & 31.25 \\
1.50 & 11693 & 7386 & 9261 & 79.75 & 100.00       & 921 & 1601 & 4796 & 33.38 & 51.79 \\
2.00 & 19988 & 8313 & 9261 & 89.76 & 100.00        & 1711 & 1817 & 4947 & 36.73 & 53.42 \\
 \multicolumn{11}{c}{} \\ 
 \multicolumn{11}{c}{\orl = 2, $n = 68,921$} \\ \hline
\hline
0.50 & 429 & 2219 & 48173 & 4.61 & 69.90      & - & - & - & - & - \\
0.75 & 980 & 4723 & 59832 & 7.89 & 86.81         & - & - & - & - & - \\
1.00 & 1731 & 7771 & 63556 & 12.23 & 92.22         & 68 & 951 & 10344 & 9.19 & 15.01 \\
1.50 & 4044 & 15423 & 67448 & 22.87 & 97.86       & 377 & 3267 & 22265 & 14.67 & 32.31 \\
2.00 & 6682 & 21194 & 68621 & 30.89 & 99.56        & 696 & 4198 & 23192 & 18.10 & 33.65 \\ 
 \multicolumn{11}{c}{} \\ 
 \multicolumn{11}{c}{\orl = 3, $n = 531,441$} \\ \hline
\hline
0.50 & 185 & 2041 & 180748 & 1.13 & 34.01         & - & - & - & - & - \\
0.75 & 380 & 5052 & 251389 & 2.01 & 47.30        & - & - & - & - & - \\
1.00 & 674 & 8328 & 291311 & 2.86 & 54.82      & 29 & 1439 & 41962 & 3.43 & 7.90 \\
1.50 & 1517 & 19700 & 375706 & 5.24 & 70.70       & 175 & 5555 & 101790 & 5.46 & 19.15 \\
2.00 & 2551 & 28360 & 415901 & 6.82 & 78.26        & 286 & 6868 & 107001 & 6.42 & 20.13 \\ 
 \multicolumn{11}{c}{} \\ 
 \multicolumn{11}{c}{\orl = 4, $n = 4,173,281$} \\ \hline
\hline
0.50 & 88 & 2372 & 645384 & 0.37 & 15.46      & - & - & - & - & - \\
0.75 & 188 & 6629 & 957724 & 0.69 & 22.95        & - & - & - & - & - \\
1.00 & 288 & 10740 & 1152954 & 0.93 & 27.63        & 17 & 2320 & 167433 & 1.39 & 4.01 \\
1.50 & 669 & 27814 & 1621138 &  2.03 &  38.85   & 84 & 9249 & 433255 & 2.13 & 10.38 \\
2.00 & 1129 & 37769 & 1856458 & 2.03 & 44.48       & 145 & 10823 & 460370 & 2.35 & 11.03 \\
\end{tabular}\label{tab:false_conn_percentage}
\end{table}

Table~\ref{table:octree_fc} provides measurements of local/cell-based false connections.
However, false connections can also have an influence on a global/network-scale connectivity. 
It is possible that a series of local-false connections can combine and create a pathway between the inflow and outflow boundaries of fracture-cells through the UDFM domain where the underlying network does not connect, cf., the bottom pathway created in the coarse mesh example provided in Figure~\ref{fig:false_connections}.
Once the UDFM mesh is created, we query the connectivity of the mesh and whether the fracture cells percolate through the domain, i.e., if there is a set of fracture cells in the UDFM mesh that connects the inflow and outflow boundaries. 
A comparison of the UDFM mesh and DFN percolation is provided in Table~\ref{tbl:percolation}.
For the percolating networks, $p^\prime \geq 1$, the topological properties match for all values of \textit{orl}.
However, for the non-percolating networks, \pp=0.5, 0.75, the topology of the UDFM only matches the DFN for the highest level of refinement \orl $ = 4$.

\begin{table}[!htb]
    \caption{Comparison of the UDFM mesh and DFN percolation. Legend: "+" indicates the UDFM percolates, "-" indicates the UDFM mesh does not percolate. Green - UDFM mesh topology coincides with the DFN topology for percolating networks. Yellow - UDFM mesh topology coincides with the DFN topology for non-percolating networks.  Red - UDFM mesh topology \textbf{does not} coincide with the DFN topology.         
    \label{tab:percolation}} \label{tbl:percolation}
    \begin{minipage}{0.35\textwidth}
    \centering

    \begin{tabular}{|c|c|c|c|c|}
    \multicolumn{5}{c}{Isolated Fracture Retained} \\ \hline
         \pp & \orl = 1 & \orl = 2 & \orl = 3 & \orl = 4\\\hline

         0.5  & \cellcolor{red!25} + & \cellcolor{red!25} + & \cellcolor{red!25} +  & \cellcolor{yellow!25} -  \\\hline
         0.75 & \cellcolor{red!25} + & \cellcolor{red!25} + & \cellcolor{red!25} +  & \cellcolor{yellow!25} -  \\\hline
         1.0  & \cellcolor{green!25} +  & \cellcolor{green!25} +  & \cellcolor{green!25}+ & \cellcolor{green!25} + \\\hline
         1.5  & \cellcolor{green!25} +  & \cellcolor{green!25} +  & \cellcolor{green!25}+ & \cellcolor{green!25} +\\\hline
         2.0  & \cellcolor{green!25} +  & \cellcolor{green!25} +  & \cellcolor{green!25}+ & \cellcolor{green!25} +\\\hline
        \end{tabular}
    \end{minipage}
    \hspace{2.2cm}
    \begin{minipage}{0.35\textwidth}
        \centering
        \begin{tabular}{|c|c|c|c|c|}
        \multicolumn{5}{c}{Isolated Fracture Removed} \\ \hline
         \pp & \orl = 1 & \orl = 2 & \orl = 3 & \orl = 4\\\hline
             0.5  &  &  & &  \\\hline
             0.75 &  &   &  &  \\\hline
             1.0  & \cellcolor{green!25} +  & \cellcolor{green!25} +  & \cellcolor{green!25}+ & \cellcolor{green!25} + \\\hline
             1.5  & \cellcolor{green!25} +  & \cellcolor{green!25} +  & \cellcolor{green!25}+ & \cellcolor{green!25} +\\\hline
             2.0  & \cellcolor{green!25} +  & \cellcolor{green!25} +  & \cellcolor{green!25}+ & \cellcolor{green!25} +\\\hline
        \end{tabular}
    \end{minipage}%
\end{table}

\subsubsection{Upscaling/Effective Hydraulic properties}
Once the domain is meshed, the upscaled/effective hydraulic properties, permeability and porosity, of each cell need to be determined. 
In the UDFM model, we take the spectral radius of a full rank dyadic permeability tensor constructed using the linear superposition of the fracture apertures, porosities, and coordinate transformation matrix to be the equivalent permeability of the Voronoi cell. 
Permeability and porosity at each fracture cell is based on the volume of fractures in the control volume, which is calculated using the intersection area of each fracture within each control volume, as well as the fracture orientations, defined by the components of their normal vectors.
The aperture of the fractures ($b_f$, [m]) are given by equation~\eqref{eq:aperture}, the surface area of a fracture within a control volume ($a_f$, [m$^2$]) is directly measured using computational geometry, and the volume of a single fracture ($v_f$, [m$^3$])  is given by their product
\begin{equation}
v_f = a_f \cdot b_f\,.
\end{equation}
Dividing $v_f$ by the total cell volume ($v_c$, [m$^3$]) provides the cell-based fracture porosity ($\phi^{f}$, [-])
\begin{equation}
\phi_f  = \frac{v_f}{v_c}\,.
\end{equation}
The components of the normal vector of the fracture ${\bf n_f} = \{n_1, n_2, n_3\}$ are  used to define a coordinate transformation tensor for that particular fracture
\begin{equation}
    \mathcal{A}_f=
        \begin{bmatrix}
            (n_2)^2+(n_3)^2 & -n_1n_2 & -n_3n_1 \\
            -n_1n_2 & (n_3)^2+(n_1)^2 & -n_2n_3 \\
            -n_3n_1 & -n_2n_3 & (n_1)^2+(n_2)^2,
        \end{bmatrix}
\end{equation}
Then equivalent fracture permeability tensor (${\bf K}_F$, [m$^2$]) is defined at each control volume in the continuum mesh by linear superposition of these terms taken over all fractures intersecting with that cell
\begin{equation}
    {\bf K}_F=\frac{1}{12}\sum_{f=1}^N \phi_f \mathcal{A}_f b_f^2\,.
\end{equation}
Notice that ${\bf K}_F$ is a 3x3 full rank second-order (dyadic) tensor and the formulation is loosely based on the cubic law.
Given the inability of most flow and transport codes to handle an anisotropic permeability tensor on unstructured meshes, we take the maximum eigenvalue ($\lambda$) of ${\bf K}_F$ to be a single equivalent fracture permeability for the cell,
\begin{equation}
K_F = \rho({\bf K}_F),
\end{equation}
where $\rho({\bf K}_F)$ denotes the spectral radius of ${\bf K}_F$ defined as 
\begin{equation}
\rho({\bf K}_F) = \max\{ |\lambda_1|, |\lambda_2|, |\lambda_3|\}\,.
\end{equation}
Likewise, the total volume $v_F$ of $N$ intersecting fractures in the control volume is the summation of individual volumes 
\begin{equation}
    v_F=\sum_{f=1}^{N}v_f=\sum_{f=1}^N a_f \cdot b_f\,,
\end{equation}
which is used to determine the equivalent porosity of a fracture cell $(\phi_F$, [-])
\begin{equation}
    \phi_F =\frac{v_F}{v_c}\,.
\end{equation}
Finally, the equivalent permeability of the fracture cell is the weighted arithmetic average of the matrix permeability $k_m$ fracture and 
\begin{equation}
    k = (1 - \phi_F) \cdot k_m + k_F,
\end{equation}
where the equivalent porosity is given by $\phi_F$. Equations (15) and (16) ensure we model the correct volumetric flux out of each cell, while limiting the transport to the fractures.~\citet{sweeney2020upscaled} verified this approach is conceptually consistent with analytical solutions.
We consider a matrix porosity value of $\phi_m = 0.01$.
These values fall within observed ranges of crystalline rock, of which our fracture network generation parameters are also loosely based~\citep{hyman2019matrix,zhou2007field}.
The matrix cells in the mesh are directly assigned values of $k_m$ and $\phi_m$.  

\subsection{Flow and transport simulations}
Once the single continuum mesh with hydraulic parameters is generated, flow and transport are then simulated.
We use the massively parallel flow and reactive code {\sc pflotran} to perform the simulations~\citep{pflotran-user-ref}.
We simulate solute transport through steady isothermal flow within the domain.
We drive flow through the domain using a pressure difference of 0.001 MPa across the $x$ axis.
Neumann, no-flow, boundary conditions are applied along lateral boundaries and gravity is not included in these simulations.
The distribution of pressure is modeled using Darcy's Law
\begin{equation}\label{eq:darcy}
{\bf q} = -\frac{k}{\mu} \nabla P\,
\end{equation}
where ${\bf q}$ is the Darcy flux [m/s], $k$ is the permeability [m$^2$], $\mu$ is the fluid viscosity [Pa s], and $P$ is the fluid pressure [Pa].
Simulations are performed with the fluid temperature at 20$^{\circ}$ C, which corresponds to a viscosity of $\mu = 8.9\mathrm{e}{-4}$ Pa s for the pressure values considered.
Once steady conditions are obtained, we can invert \eqref{eq:darcy} and obtain a numerical estimation for the effective permeability of the block
\begin{equation}\label{eq:k_eff}
k_{\text{eff}} = -\mu\frac{{\bf q}}{\nabla P}\,.
\end{equation}
The values on the right hand side of $k_{\text{eff}}$ are directly observable from the simulations. 

We model solute transport using an Eulerian formulation. 
We consider conservative, decaying, and adsorbing tracers.
The movement of a conservative (non-reactive) tracer with concentration $C(t,\vx)$ is governed by 
\begin{equation}\label{eq:ade_conservative}
\frac{\partial}{\partial t} \varphi C + {\boldsymbol{\nabla}}\cdot\Big({\boldsymbol{q}}C - \varphi D {\boldsymbol{\nabla}}(C)\Big) = 0,
\end{equation}
where $D$ is the dispersion coefficient that is defined here solely by the molecular diffusion coefficient and is set to $10^{-9}$ $\textrm{m}^2$/s.
The movement of the decaying tracer ($\varphi C^\ast$) is governed by 
\begin{equation}\label{eq:ade_decay}
\frac{\partial}{\partial t} \varphi C^\ast + {\boldsymbol{\nabla}}\cdot\Big({\boldsymbol{q}}C^\ast - \varphi D {\boldsymbol{\nabla}}(C^\ast )\Big) = - \lambda \varphi C^\ast,
\end{equation}
where $\lambda$ is the decay rate constant [1/s], which is used to impose a half life of 100 years. 
The movement of the adsorbing tracer ($\varphi C^\prime$) with retardation coefficient ($R$, [-]) is governed by
\begin{equation}\label{eq:ade_adsorbing}
\frac{\partial}{\partial t} \varphi C^\prime + \frac{1}{R}{\boldsymbol{\nabla}}\cdot\Big({\boldsymbol{q}}C^\prime  - \varphi D {\boldsymbol{\nabla}}(C^\prime )\Big) = 0.
\end{equation}
The retardation coefficient $R$ is expressed as
\begin{equation}
R = 1 + \frac{1}{\varphi s_{l}\rho_w}K^D,
\end{equation}
where $s_{l}$ is liquid saturation [-], $\rho_{w}$ is water density [kg/m$^3$], and $K^{D}$ is the distribution coefficient [kg/m$^3$]. The porosity will vary across the domain based on the upscaled porosity values. The distribution coefficient represents the ration of sorbed to aqueous concentrations. Here, we set the distribution coefficient to $R = 4 \cdot 10^3$.


All tracers are subject to the following initial and boundary conditions. 
Initially, there is no tracer in the domain ($\Omega$)
\begin{equation}\label{eq:ade_ic}
C(0,\vx) = 0 \qquad \forall \qquad  \vx \in \Omega.
\end{equation}
Tracer is introduced uniformly into the domain across the entire inlet face ($\Gamma_I$) using a pulse injection
\begin{equation}\label{eq:ade_bc}
C(0,\vx) = C_0 \delta(0,\vx) \qquad \forall \qquad  \vx \in \Gamma_I\,,
\end{equation}
where $\delta(0,\vx)$ is the Dirac delta function applied to the inlet face at time $t=0$.
The inflow boundary is assigned a zero-gradient boundary condition to prevent solute diffusing out of the domain in that direction. 
The outflow boundary ($\Gamma_O$) is assigned an adsorbing boundary condition.
All other boundaries are reflective.
Transport simulations are run for 10$^8$ years. 

We measure the tracer mass flow (mol/yr) through the outflow boundary as a function of time, which we refer to as the breakthrough curve (BTC).
We normalize these measurements by the initial tracer mass (mol) injected into the system.
Time is normalized by the peak arrival time of the BTC for \orl = 1 for the conservative tracer.
These normalizations facilitate easier comparisons between tracer types (conservative, decaying, and sorbing), \orl values, fracture densities, and whether isolated fractures are retained or removed.

\section{Results}
\label{sec:results}
In this section, we begin by reporting the measurement of effective permeability. 
Next, we discuss the breakthrough curves  obtained in our ensemble of networks. 

\subsection{Effective permeability}

\begin{figure}[htb!]
     \centering
     \subfloat[]{\includegraphics[width=0.54\textwidth]{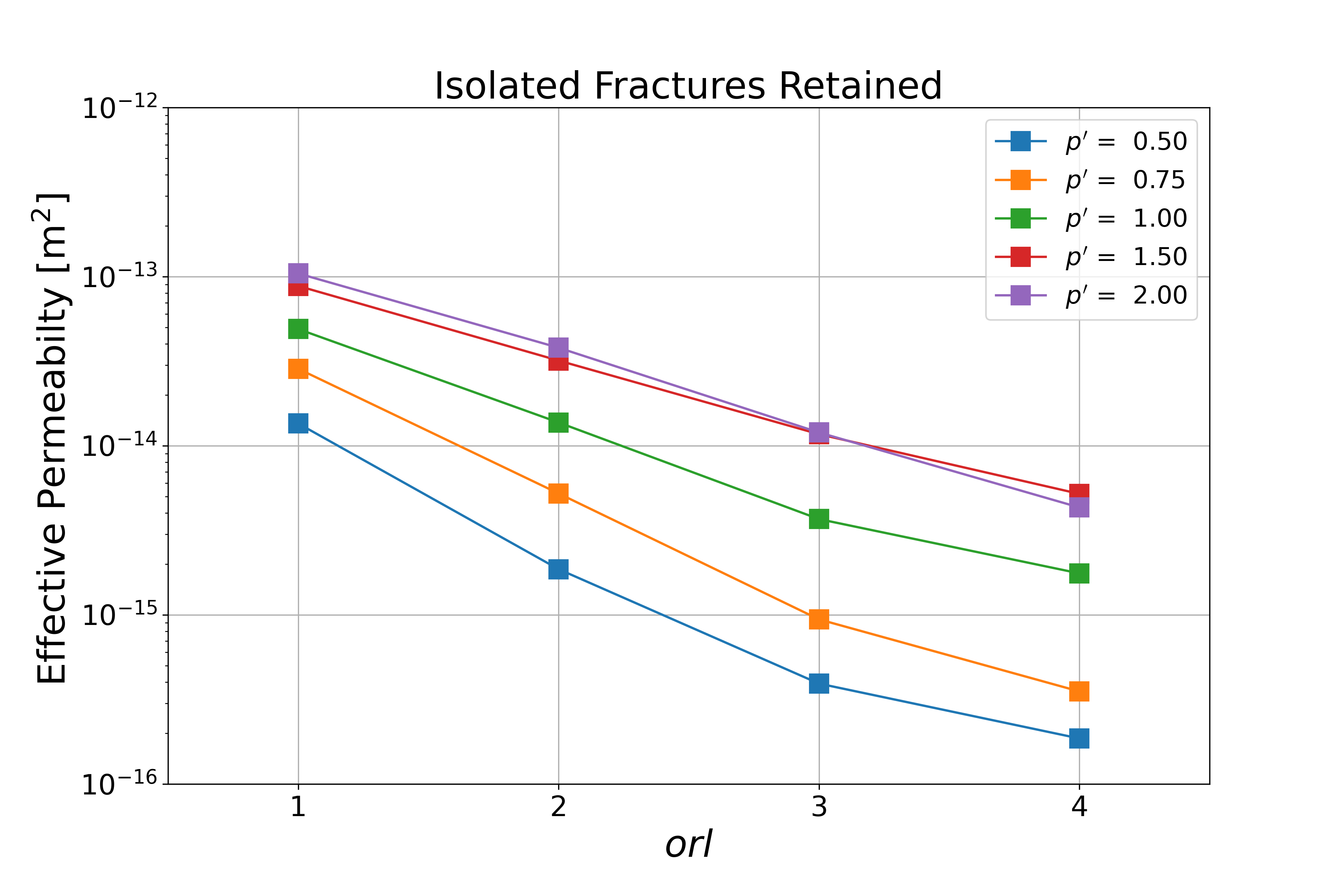}}
     \subfloat[]{\includegraphics[width=0.54\textwidth]{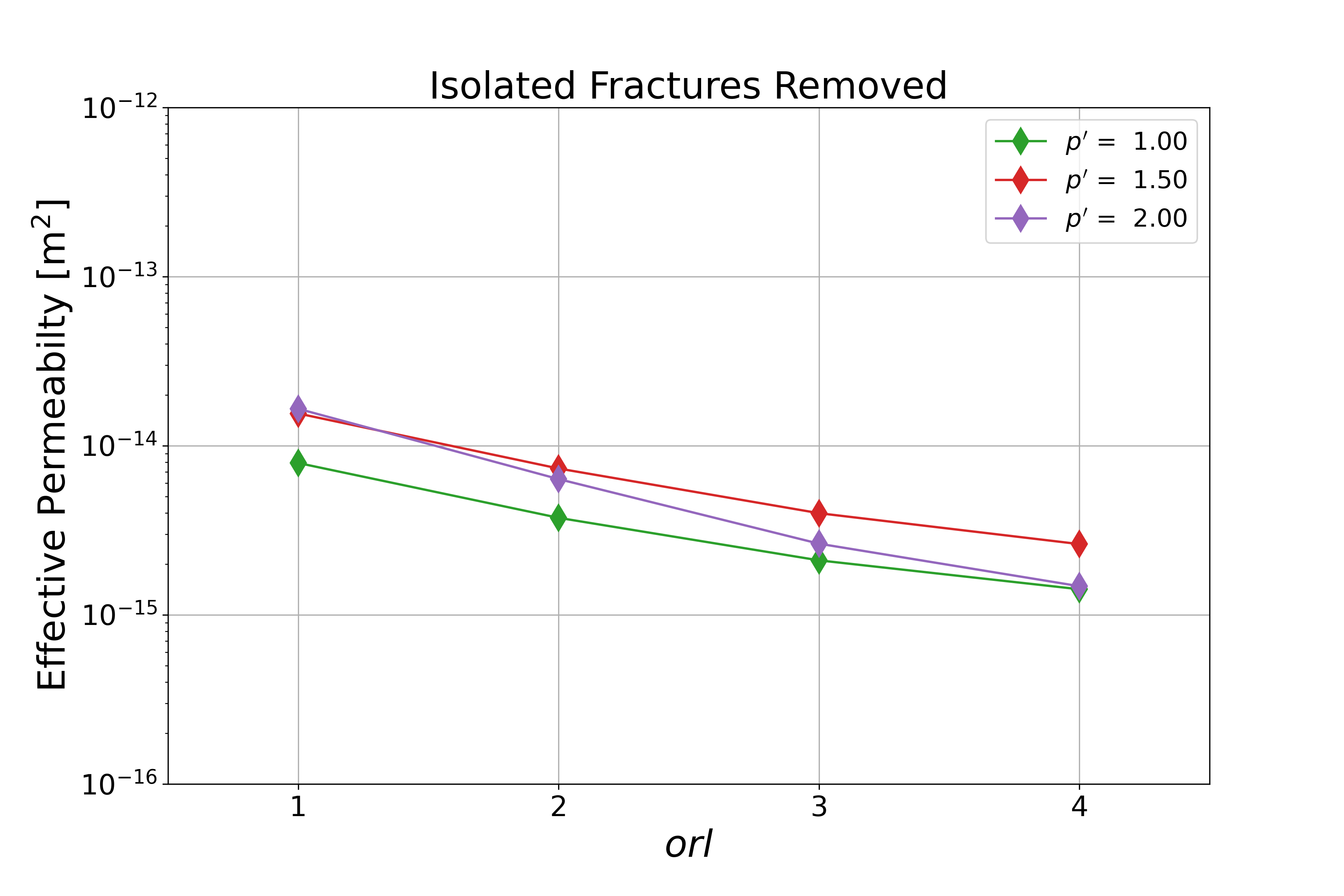}} \\
     \subfloat[]{\includegraphics[width=0.54\textwidth]{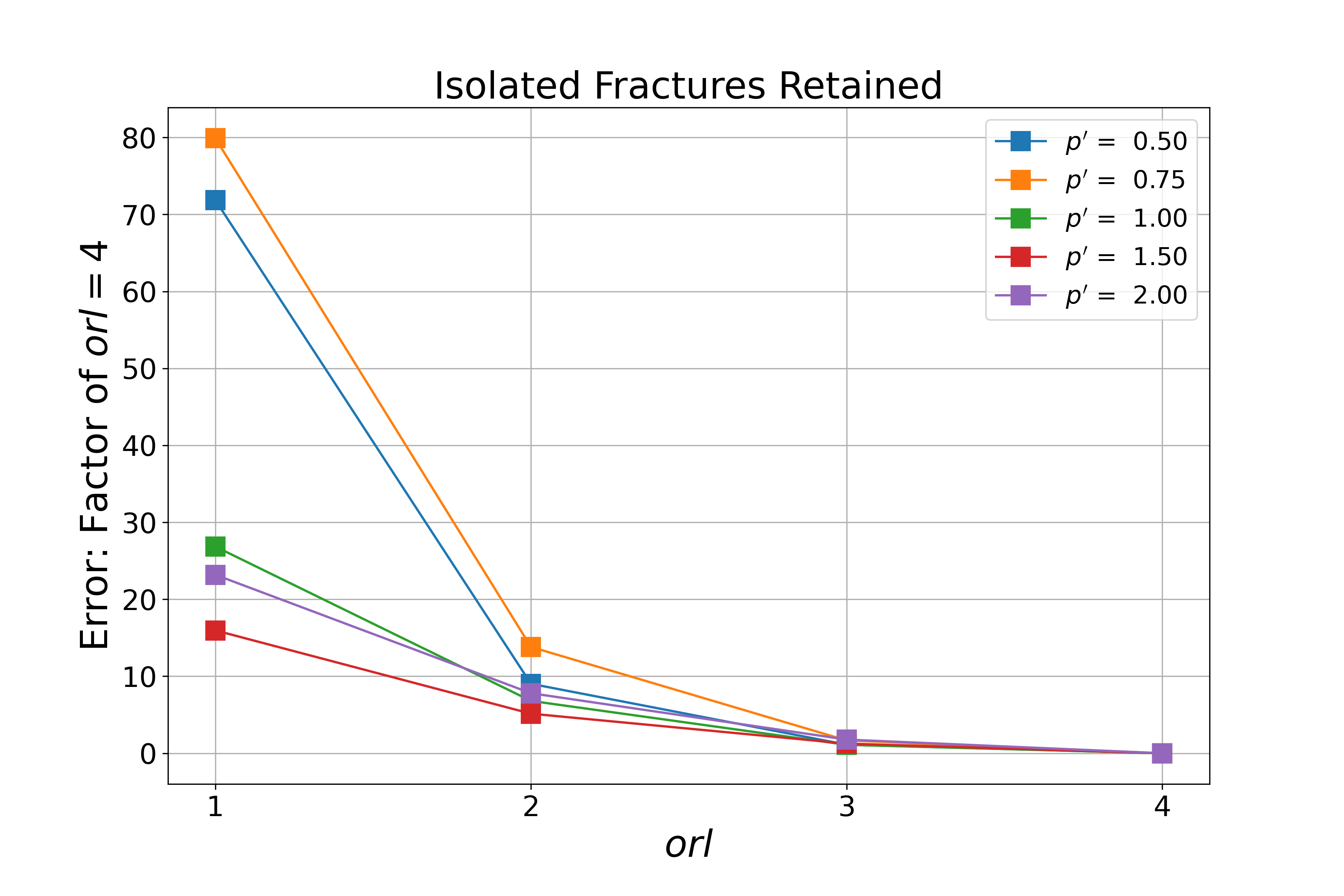}}
     \subfloat[]{\includegraphics[width=0.54\textwidth]{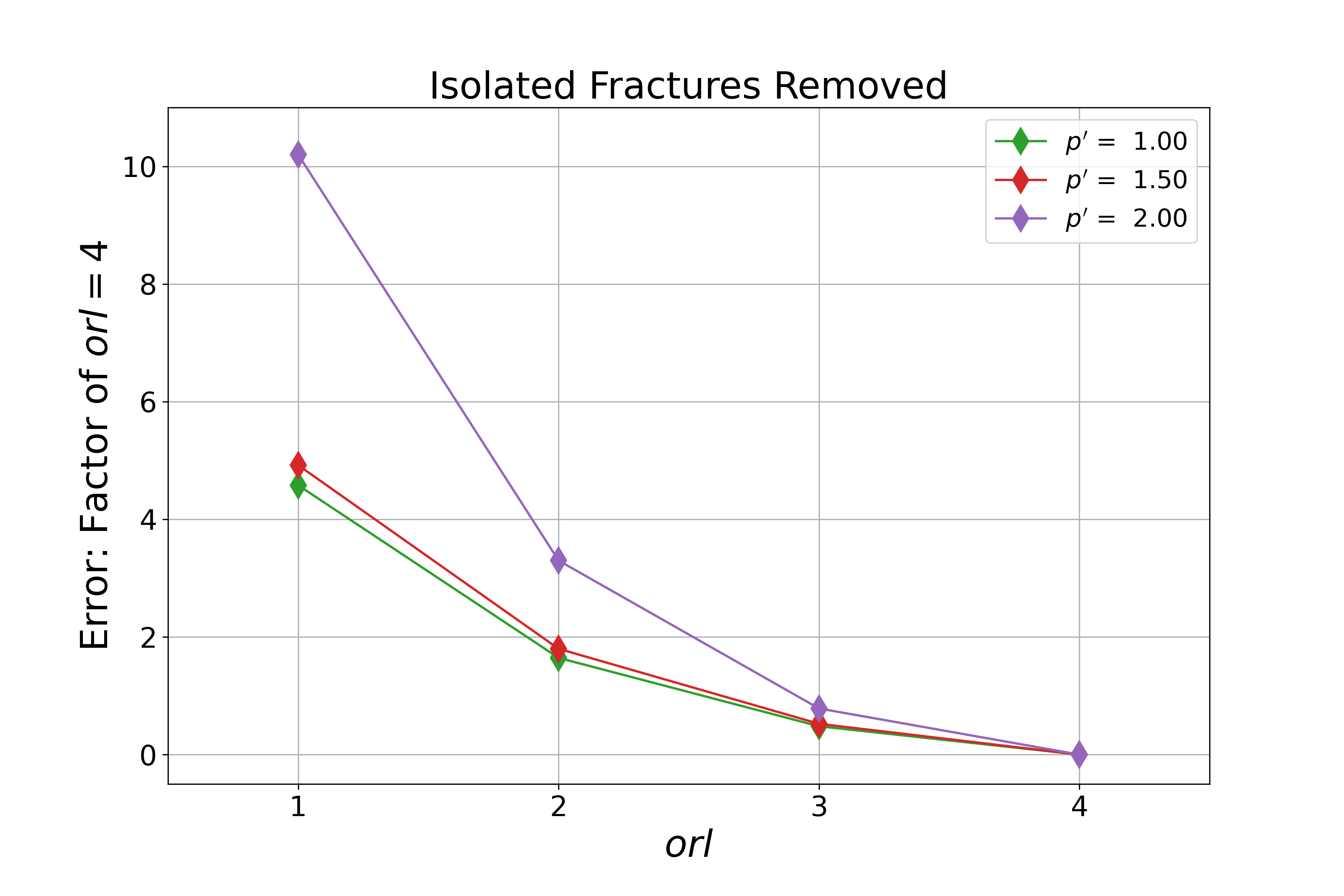}}\\
     \caption{Effective permeability of samples at different \orl values with a matrix permeability of $1\cdot 10^{-16}$ m$^2$. (a)  Isolated Fractures Retained -  Effective Permeability  (b) Isolated Fractures Removed - Effective Permeability  (c) Isolated Fractures Retained  -  Error factor of effective permeability. (d) Isolated Fractures Removed  -  Relative error of effective permeability.}
    \label{fig:effective_perm}
\end{figure}

We populate the right hand side of \eqref{eq:k_eff} using the numerical solution to obtain an estimate for the effective permeability $k_{\text{eff}}$ of each realization.
For these samples, we considered a matrix permeability of $1\cdot 10^{-16}$ m$^2$.
The values are plotted in the top row of Figure~\ref{fig:effective_perm} with "isolated fractures retained" realizations on the left and "isolated fractures removed" on the right.
Each line corresponds to a density value $p^\prime$.
Values of $k_{\text{eff}}$ increase with increasing \pp values due to larger numbers of fractures with higher permeability being placed into the domain. 
The values of $k_{\text{eff}}$ monotonically decrease with increasing \orl values, in all cases.
Recall that the background matrix permeability is $10^{-16}$ m$^2$, which provides a lower bound to $k_{\text{eff}}$. 
The difference between $k_{\text{eff}}$ at subsequent \orl levels decreases as \orl increases, which indicates convergence to a relatively stable value should occur with a sufficient number of refinement levels.
Due to computational limitations, expanding to higher \orl to identify this value was not possible for the higher density values \pp.
Nonetheless, we examine the convergence rate by computing the relative difference/error $(e_i)$ of \orl=1,2,3 compared to the value obtained with \orl=4 using the definition
\begin{equation}
e_i = \left | \frac{ k_{\text{eff}}^i - k_{\text{eff}}^4}{k_{\text{eff}}^4} \right |,
\end{equation}
where $k_{\text{eff}}^i $ denotes the effective permeability at \orl = i. 
These values are plotted in the lower row of Figure~\ref{fig:effective_perm}.
In the case of "isolated fractures retained", the error values seem to decrease exponentially with \orl values.
The reported error for \orl=1 ranges from $\approx 15-80$ times that of \orl = 4, with the highest values observed for the non-percolating networks, \pp $<$ 1.
As we increase the number of refinement levels, the error decreases, and by \orl = 3, the estimates are all less than 2, which given the small absolute values, $\mathcal{O}(10^{-14} - 10^{-15})$, are likely acceptable.
In the case of "isolated fractures removed", the error values also declines exponentially with \textit{orl}.
ere, the estimates provided at the lower \orl levels are much closer to the \orl = 4 value when compared to the "isolated fractures retained" counter parts. 
For \orl = 3, the values are less than 1.

\begin{figure}[htb!]
     \centering
     \subfloat[]{\includegraphics[width=0.54\textwidth]{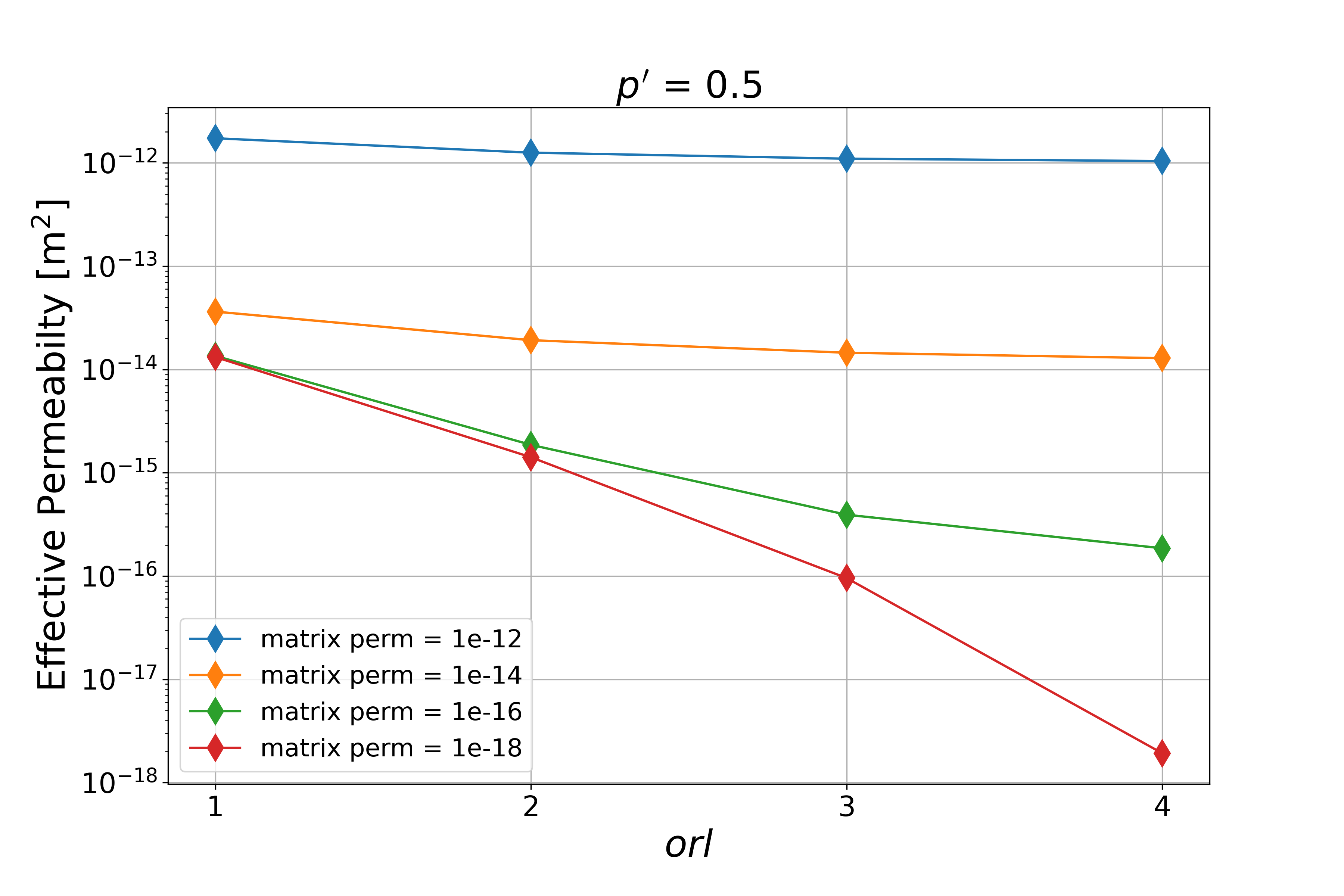}}
     \subfloat[]{\includegraphics[width=0.54\textwidth]{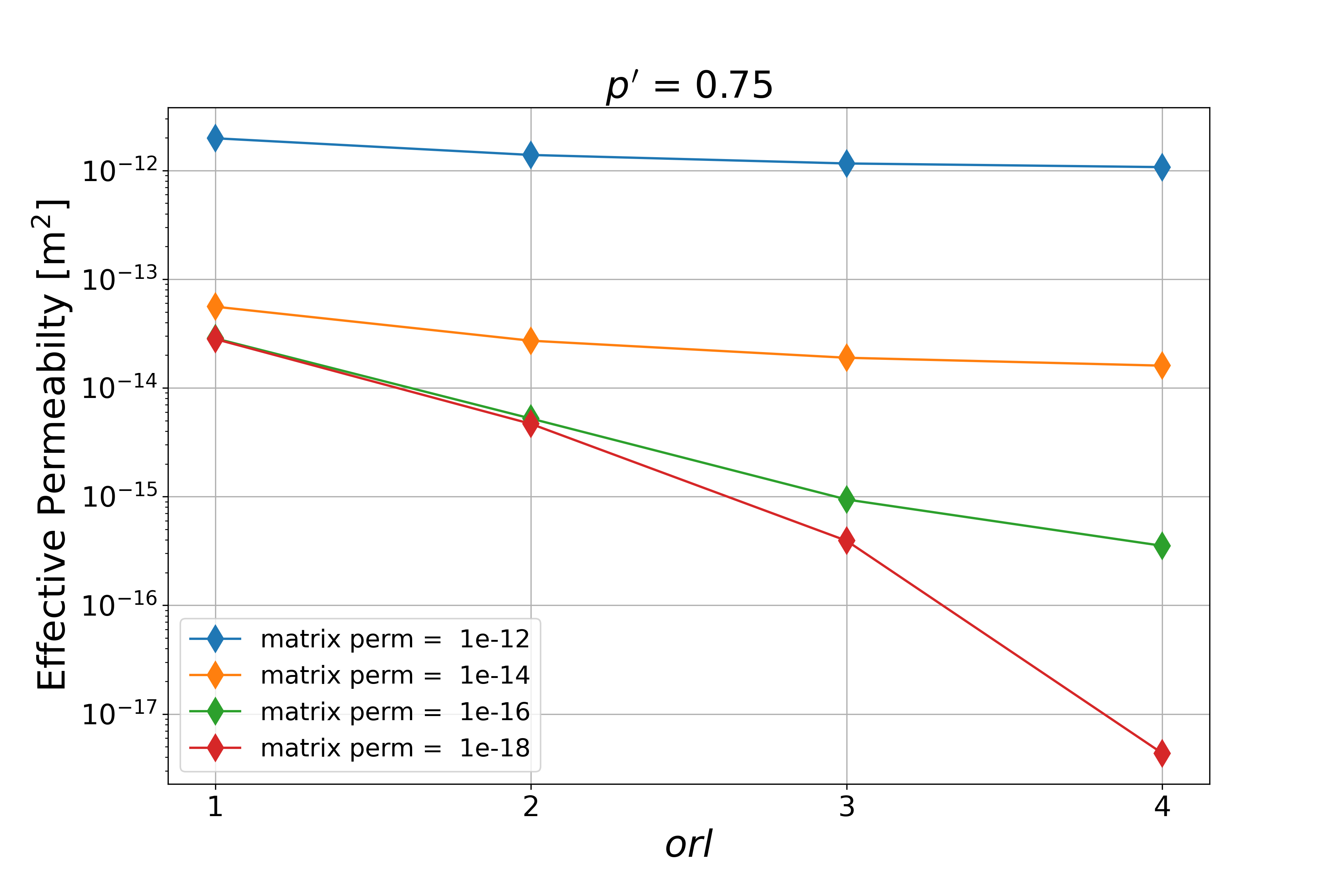}} \\
     \subfloat[]{\includegraphics[width=0.54\textwidth]{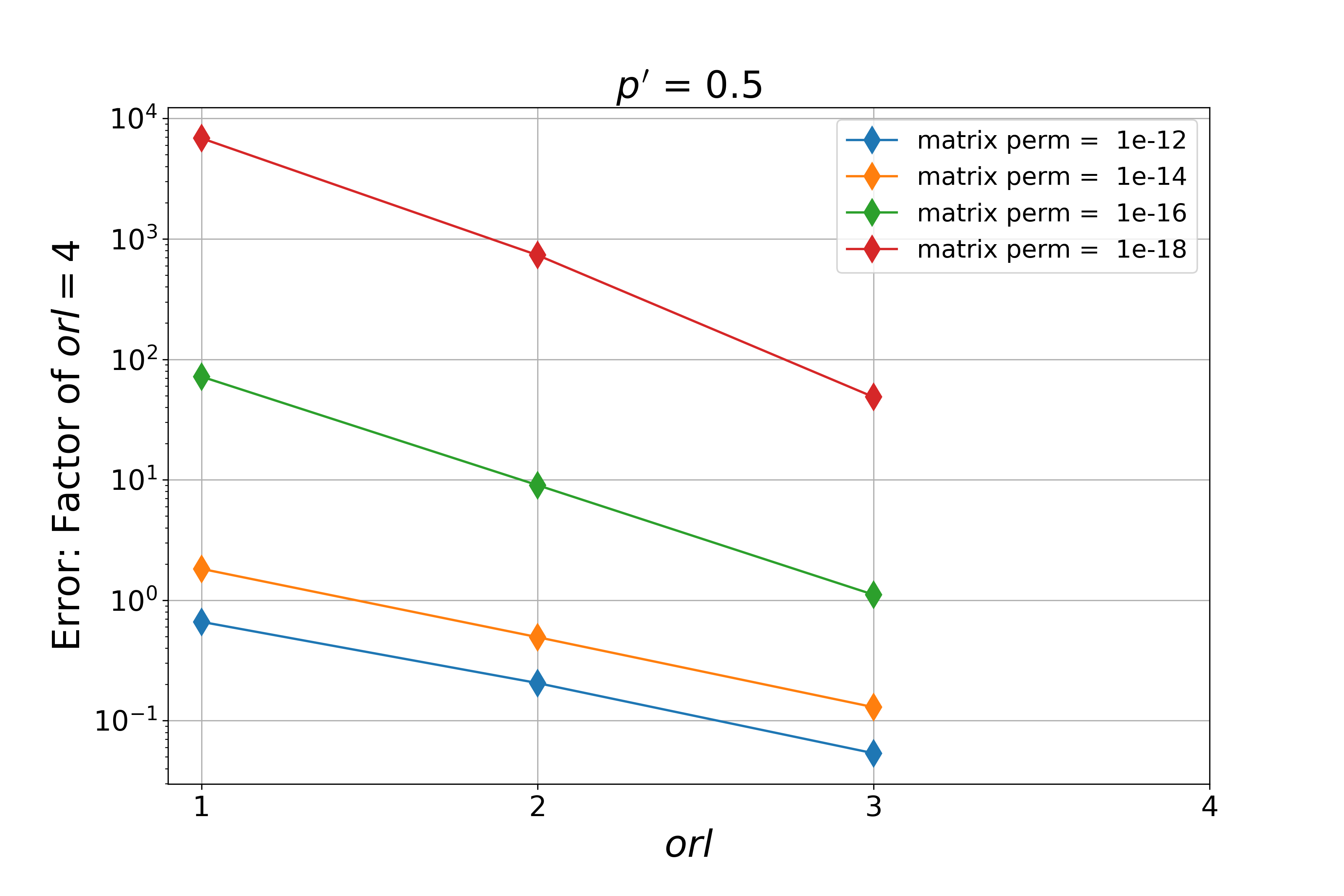}}
     \subfloat[]{\includegraphics[width=0.54\textwidth]{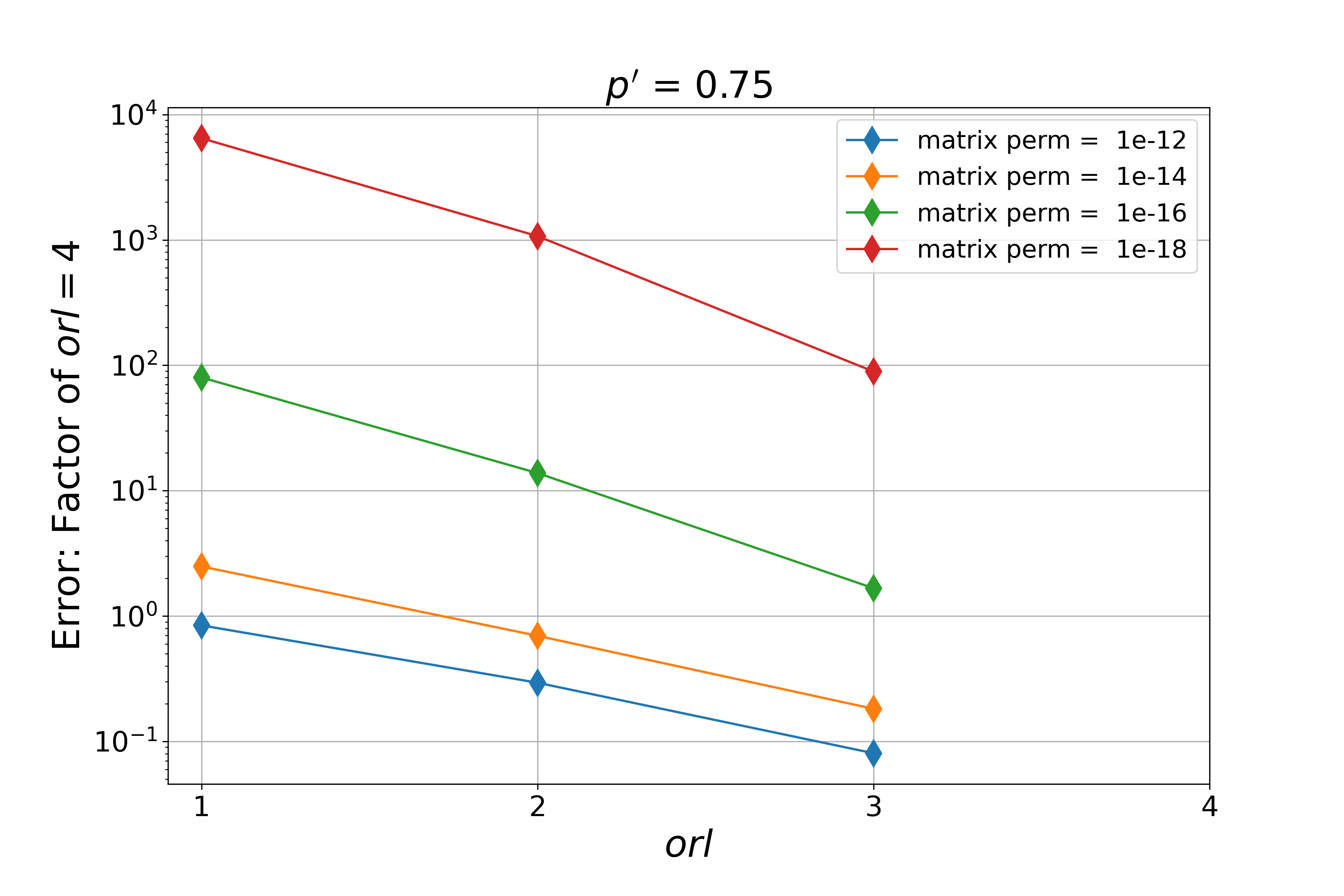}}\\
     \caption{Effective Permeability of samples at different matrix permeability values. (a) \pp = 0.5 - Effective
     Permeability. (b) \pp = 0.75 - Effective Permeability. 
     (c) \pp = 0.50 - Error factor of effective permeability.
     (d) \pp = 0.75 - Error factor of effective permeability.}
    \label{fig:matrix_effective_perm}
\end{figure}

Figure~\ref{fig:matrix_effective_perm} reports the effective permeability for additional matrix permeability values of $1\cdot 10^{-12}$ m$^2$, $1\cdot 10^{-14}$ m$^2$, $1\cdot 10^{-16}$ m$^2$, and $1\cdot 10^{-18}$ m$^2$ for \pp = 0.50 and \pp = 0.75, but with isolated fractures retained. 
We only consider the samples because they are below the percolation threshold and the topology of the UDFM mesh and DFN do not match. 
Similar to the previous samples, the effective permeabilities decrease with \orl value. 
However, the rate of decreasing, depends on the background matrix permeability. 
For the highest values, there is relatively little change with \orl value when compared to the lower values. 
A typical fracture permeability value in our model setup is around $1\cdot 10^{-9}$ m$^2$. 
So, the case with $k_m = 1\cdot 10^{-12}$ m$^2$, the domain is relatively homogeneous. 
However, for the lowest matrix permeability values, $1\cdot 10^{-18}$ m$^2$, the domain is highly heterogeneous, with permeability values ranging close to 10 orders of magnitude in the domain. 
In turn, if a macro-scale false connection forms, as in these cases, there is highly permeability channel through the domain. 
This results in effective permeability errors on the order of 4 orders of magnitude. 
Most notably, the effective permeability for the  $1\cdot 10^{-18}$ m$^2$ case does not stabilize as the values for the other ones do.

\subsection{Breakthrough curves}

The breakthrough curves (BTC) of the tracers through the outlet face of the domain as a function of time are discussed in this section. 
We begin with the two densities with networks that do not percolate (\pp = 0.5 \& 0.75).
Then we discuss the three densities that do percolate (\pp = 1.0, 1.5 \& 2.0).

\subsubsection{$p^\prime < 1$}


\begin{figure}
    \centering
    \subfloat[]{\label{subfig:p05_iso}{\includegraphics[width=0.54\columnwidth]{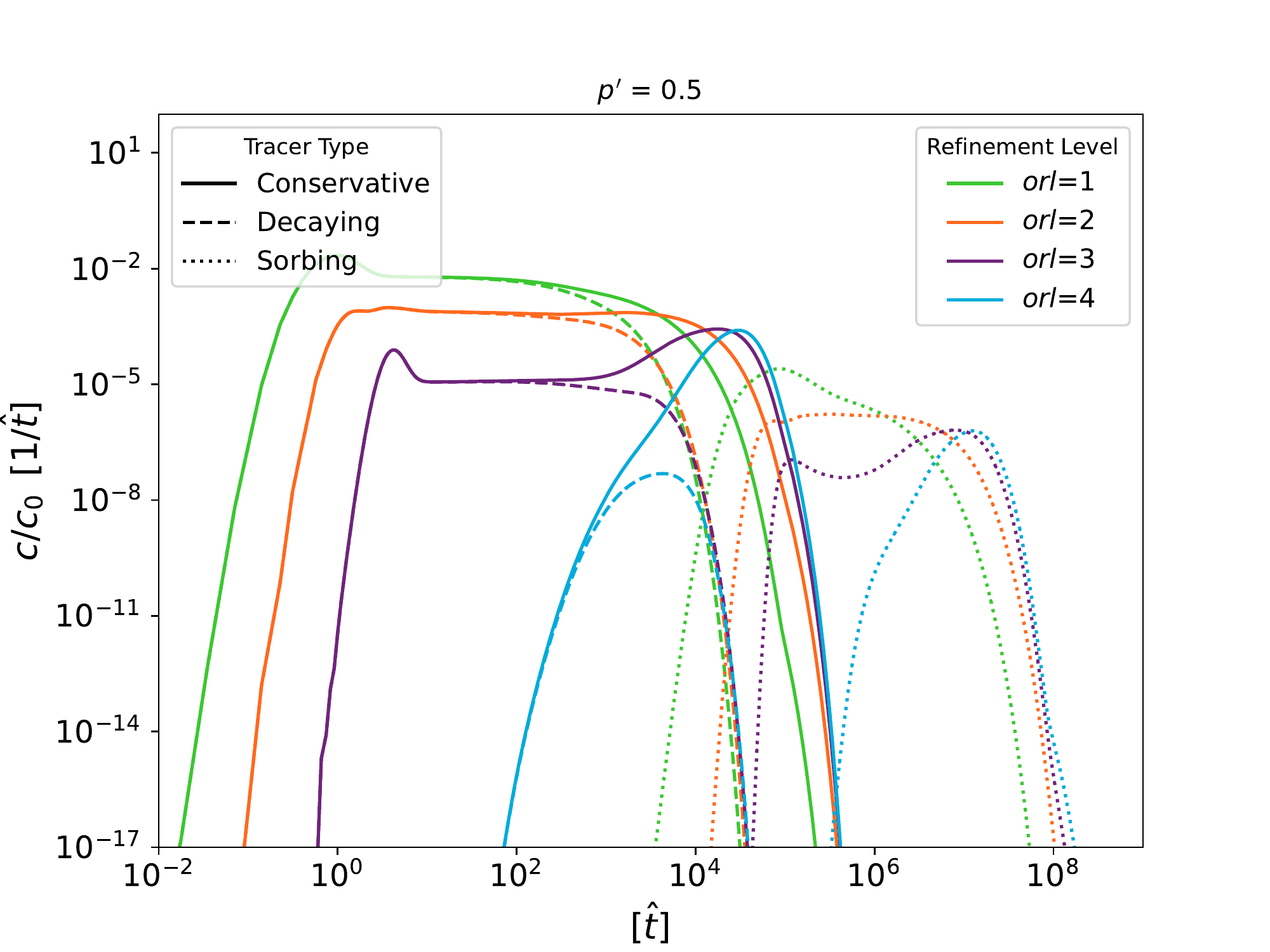}}}
    \subfloat[]{\label{subfig:p075_iso}{\includegraphics[width=0.54\columnwidth]{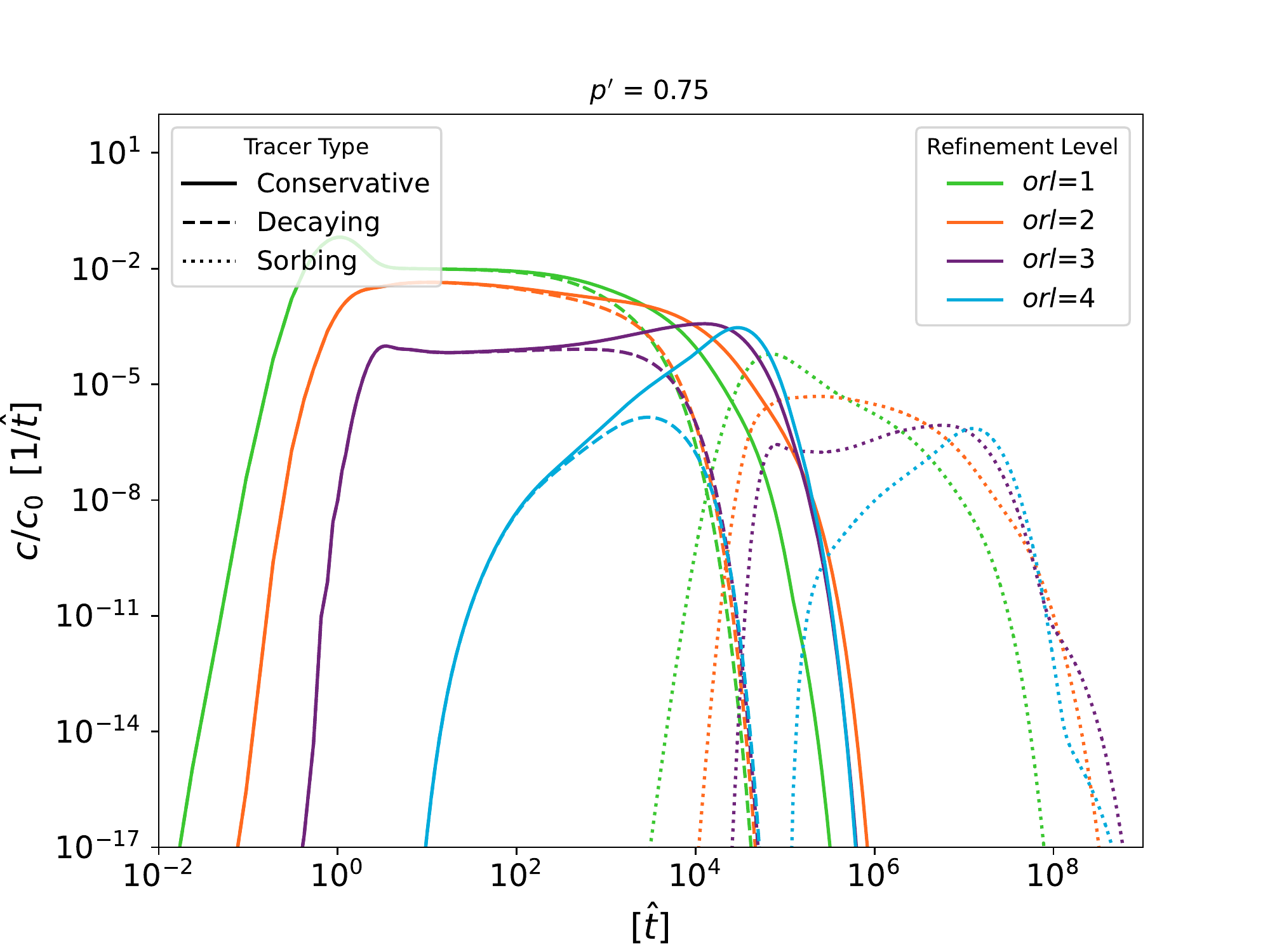}}}\\
    \caption{Isolated Fracture Retained (a) $p^\prime = 0.50$ and (b) $p^\prime = 0.75$. Solid lines are the conservative tracer. Dashed lines are the decaying tracer, and dottted lines are the sorbing tracer. Colors correspond to \orl values: 1 - green. 2 - orange, 3 - purple, 4 - blue. }
    \label{fig:p_lt_1_btc}
\end{figure}

Figure~\ref{fig:p_lt_1_btc}~(a) shows the breakthrough curves in the fracture network with density $p^\prime = 0.5$. 
We begin the descriptions with the conservative tracer, which is shown as solid lines. 
For octree refinement level \orl=1, there is an initial peak and then a plateau of tracer slowly exiting the domain.   
For octree refinement levels \orl=2 and 3, there are two distinct peaks. 
There is an initial peak corresponding to transport through the fractures and then there is a second peak corresponding to transport through the matrix. 
At \orl=4, there is only a single peak arriving at the same time as the secondary peaks in \orl=2 and 3. 
Recall that for this density, there was not only a significant number of false connections for \orl=1, 2, and 3, but the global topology of the UDFM model did not match the DFN, i.e., fracture cells in the UDFM mesh percolate through the domain while the DFN does not percolate, cf. Table~\ref{tab:percolation}. 
Thus, there is a connected pathway of fracture cells in the UDFM model for \orl=1, 2, and 3, which leads to the early peak. 
However, physically speaking, the tracer must travel through the matrix, which is the case for \orl=4. 
In the case of the decaying tracer, \orl=1, 2, and 3 also show an initial peak in tracer concentration due to transport through the fracture cells but the tracer decays before we observe the second peak, which corresponds to travel through the matrix. 
For \orl=4, there is a peak of tracer passing through the matrix, but is significantly delayed. 
The sorbing tracer breakthrough curves roughly resemble the conservative tracer curves, being significantly delayed due to adsorption, though they are more spread out than the conservative and decaying tracers.
At the exit through the matrix, the peak height of tracer concentration has reduced significantly due to delay.  

At \orl=1, there a wide spread in arrival times for all tracers.
As we increase the octree refinement level from \orl=2 to \orl=4 for the fracture network with removed isolated fractures, the actual behavior of the conservative tracer is revealed -- the initial peak in tracer concentration disappears because there is no fracture connection from the inflow to the outflow boundary. 

The breakthrough curves for a fracture network with density \pp = 0.75 is shown in Figure~\ref{fig:p_lt_1_btc}~(b).
Like \pp = 0.5, the DFN does not percolate, but \orl=1, 2 and \orl=3 UDFM meshes do. 
In general, the behavior is quite similar to what was observed for the \pp = 0.5 network. 
Once the percolation structure of the UDFM model matches the DFN, then matrix dominated transport is observed. 
The falsely percolating simulations show a dual peak structure that indicates an initial peak through a connected fracture network and a second peak due to transport through the matrix at the exit.
The profile of the BTC for the sorbing tracers are delayed and somewhat different from conservative tracer, i.e., they have less pronounced peaks.

\subsubsection{$p^\prime \geq 1$}

\begin{figure}[htb!]
    \centering
    \subfloat[]{\label{subfig:p1_iso}{\includegraphics[width=0.54\columnwidth]{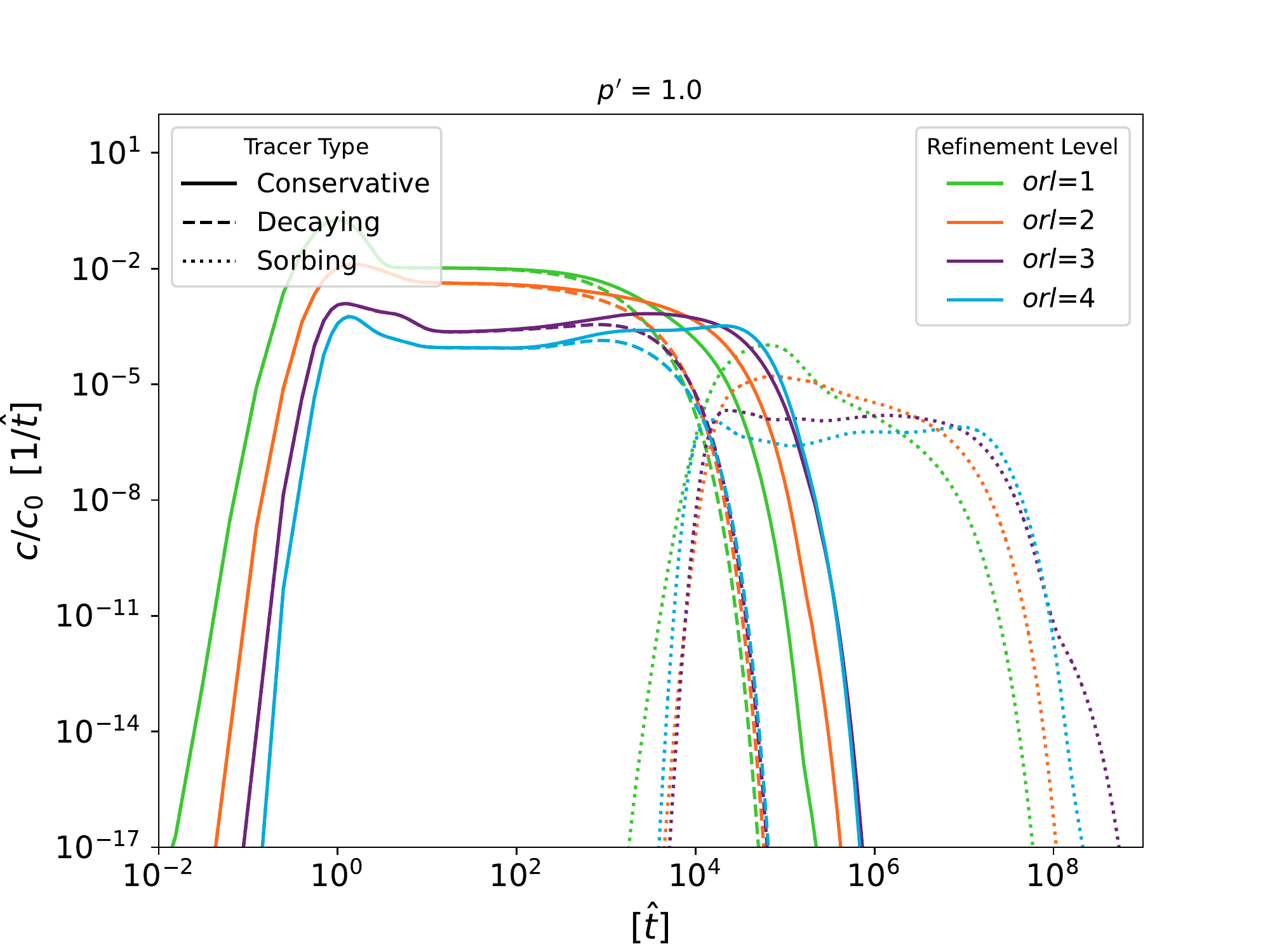}}}
    \subfloat[]{\label{subfig:p1}{\includegraphics[width=0.54\columnwidth]{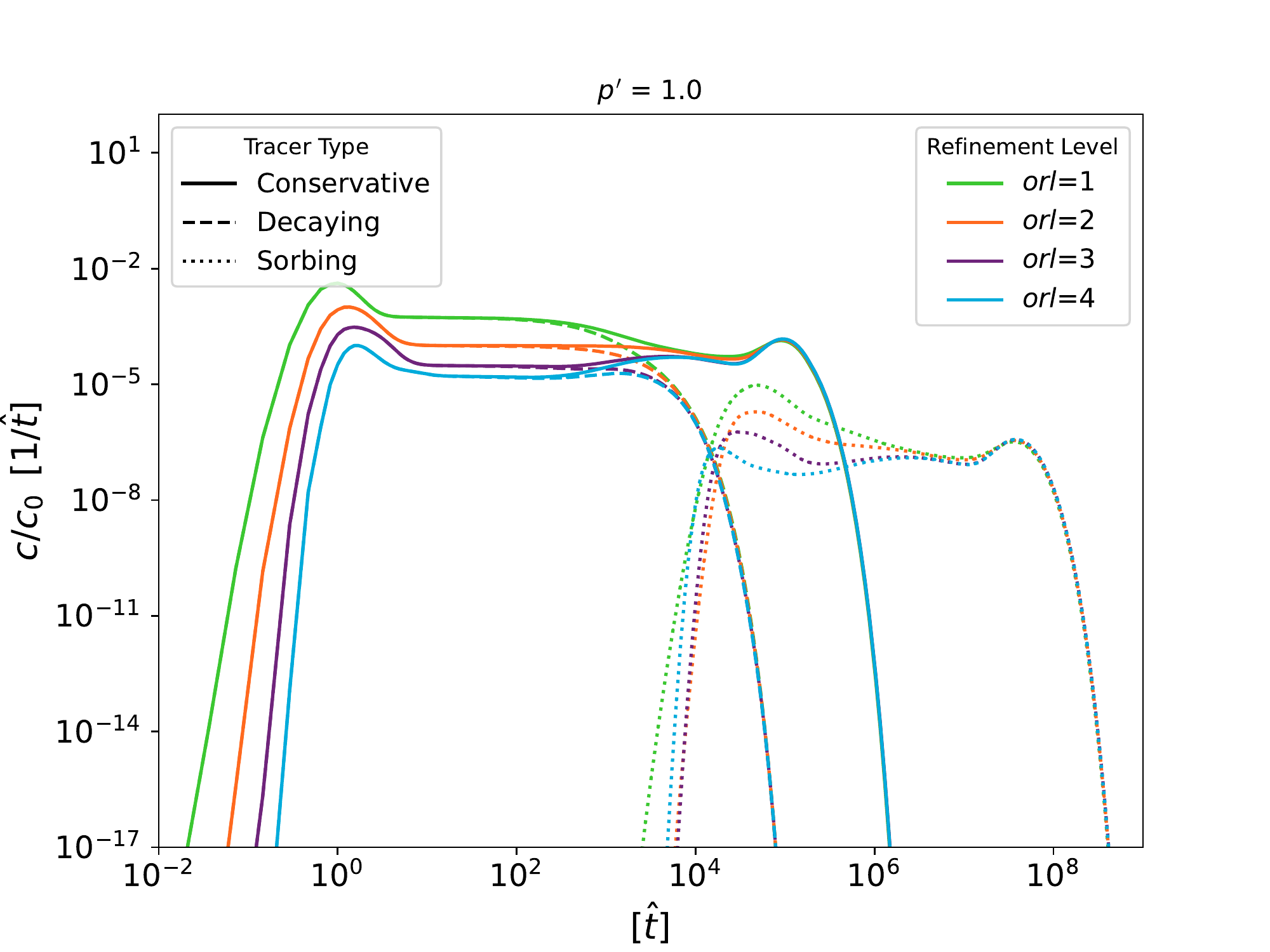}}} \\
    \caption{Breakthrough curves of Eulerian tracer in a fracture network with density $p^\prime = 1.0$. (a) isolated fractures retained and (b) isolated fractures removed.}
    \label{fig:p_1_btc}
\end{figure}

Figure~\ref{fig:p_1_btc} shows the BTCs for $p^\prime = 1.0$: (a) isolated fractures retained and (b) isolated fractures removed.
Colors and line styles are the same as in Figure~\ref{fig:p_lt_1_btc}.
Rather than seeing a stark change in the BTC behavior due to a change in global connectivity, we observe the influence of changes in local connectivity and mesh refinement. 
Although the global connectivity of the UDFM model and DFN correspond for all \orl values, the number of local false connections rapidly declines with increasing \orl value, cf. Table~\ref{table:octree_fc}.
For the case where isolated fractures are retained, 59\% of control volumes have a false connection at \orl=1, compared to less than 1\% at \orl =4. 
In the case of isolated fractures are retained and \orl=1, we see a single peak arrival with a long tail for all tracer types. 
As the \orl increases, we see the formation of a second, matrix-transport-dominated peak.
This latter peak is likely the result of better defined pathways through the system due to mesh refinement. 

For the case where isolated fractures are removed, the convergence with rising \orl is more pronounced.
This is true especially true for transport through the matrix where the tails of the distribution collapse onto one another. 
Recall that the number of local false connections is much smaller compared to the case where isolated fractures are retained. 


\begin{figure}[htb!]
    \centering
    \subfloat[]{\label{subfig:p15_iso}{\includegraphics[width=0.54\columnwidth]{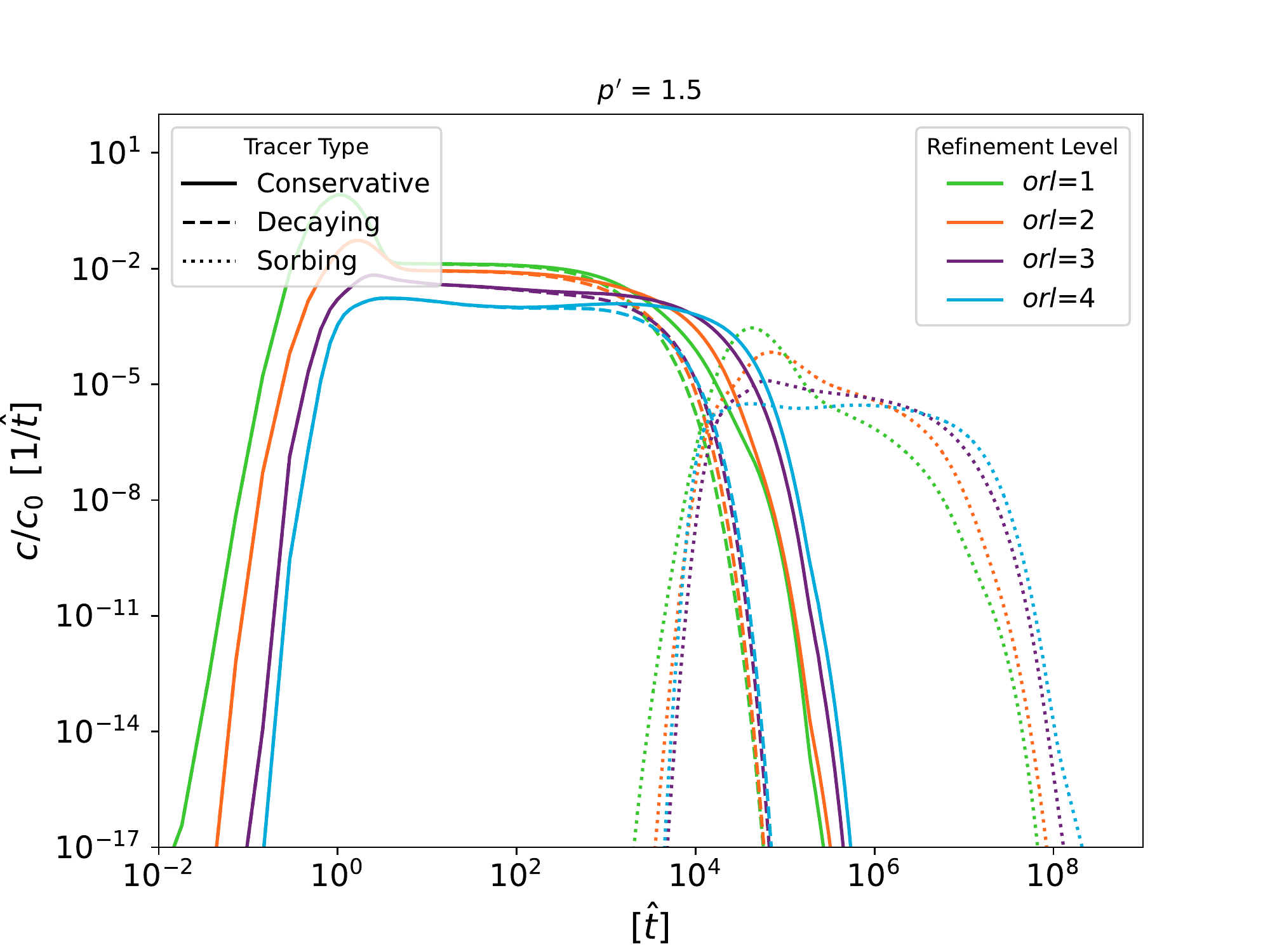}}}
    \subfloat[]{\label{subfig:p15}{\includegraphics[width=0.54\columnwidth]{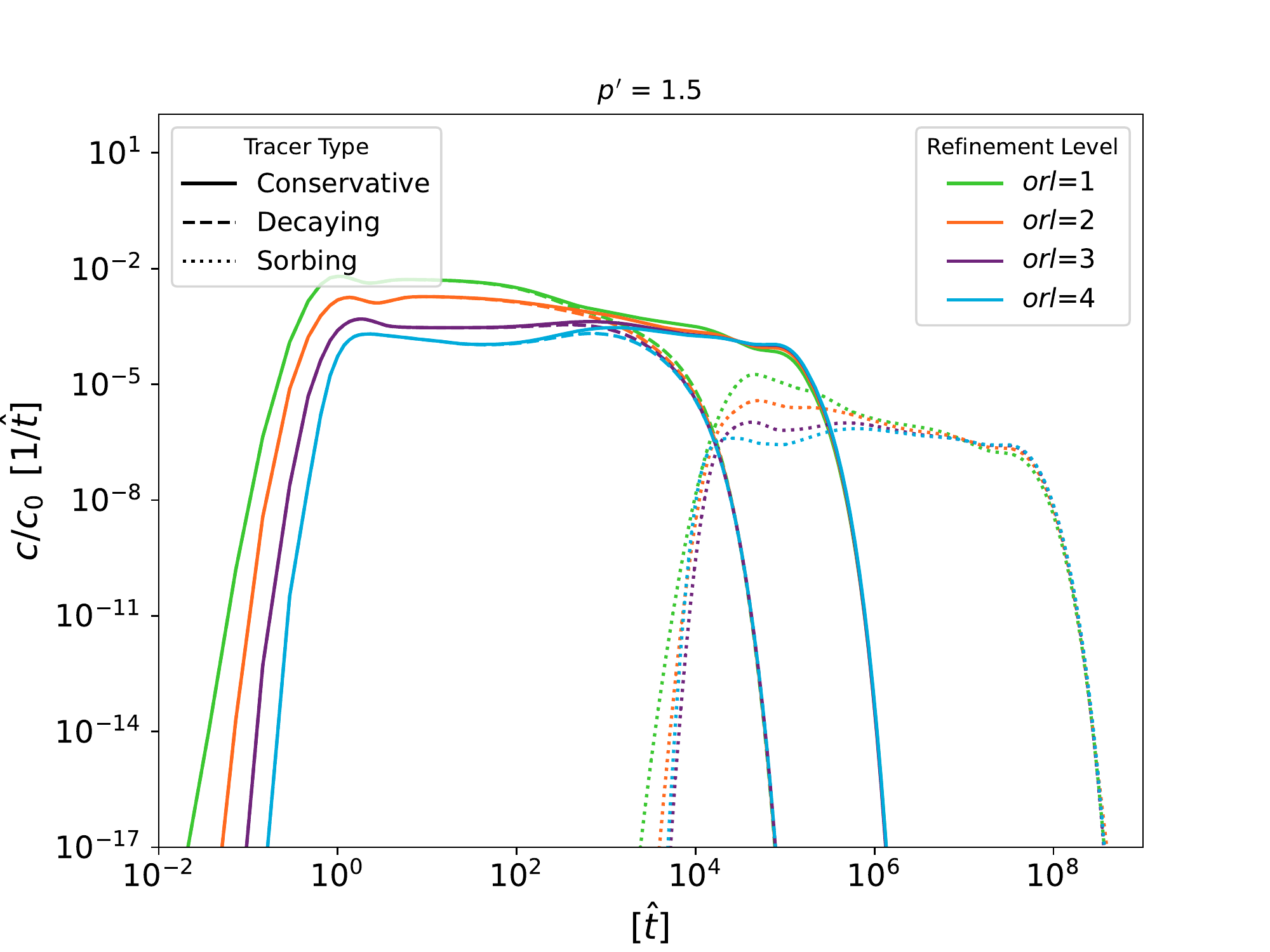}}}\\
    \caption{Breakthrough curves of Eulerian tracer in a fracture network with density $p^\prime = 1.5$. $c$ is the concentration of the tracer at time $\hat{t}$, and $c_0$ is the total amount of tracer in the pulse. We examine the solute transport for fracture networks with (a) isolated fractures retained, and (b) isolated fractures removed.}
    \label{fig:p_15_btc}
\end{figure}

Figure~\ref{fig:p_15_btc} depicts the BTCs for $p^\prime = 1.5$.
The BTC show more or less the same behavior as in Figure~\ref{fig:p_1_btc} with respect to how the BTC is influenced by the \orl value and whether or not isolated fractures are retained or removed.
A small difference is that the latter peak, which corresponds to matrix transport, does not show the same collapse, nor the same valley between the two peaks. 

\begin{figure}[htb!]
    \centering
    \subfloat[]{\label{subfig:p2_iso}{\includegraphics[width=0.54\columnwidth]{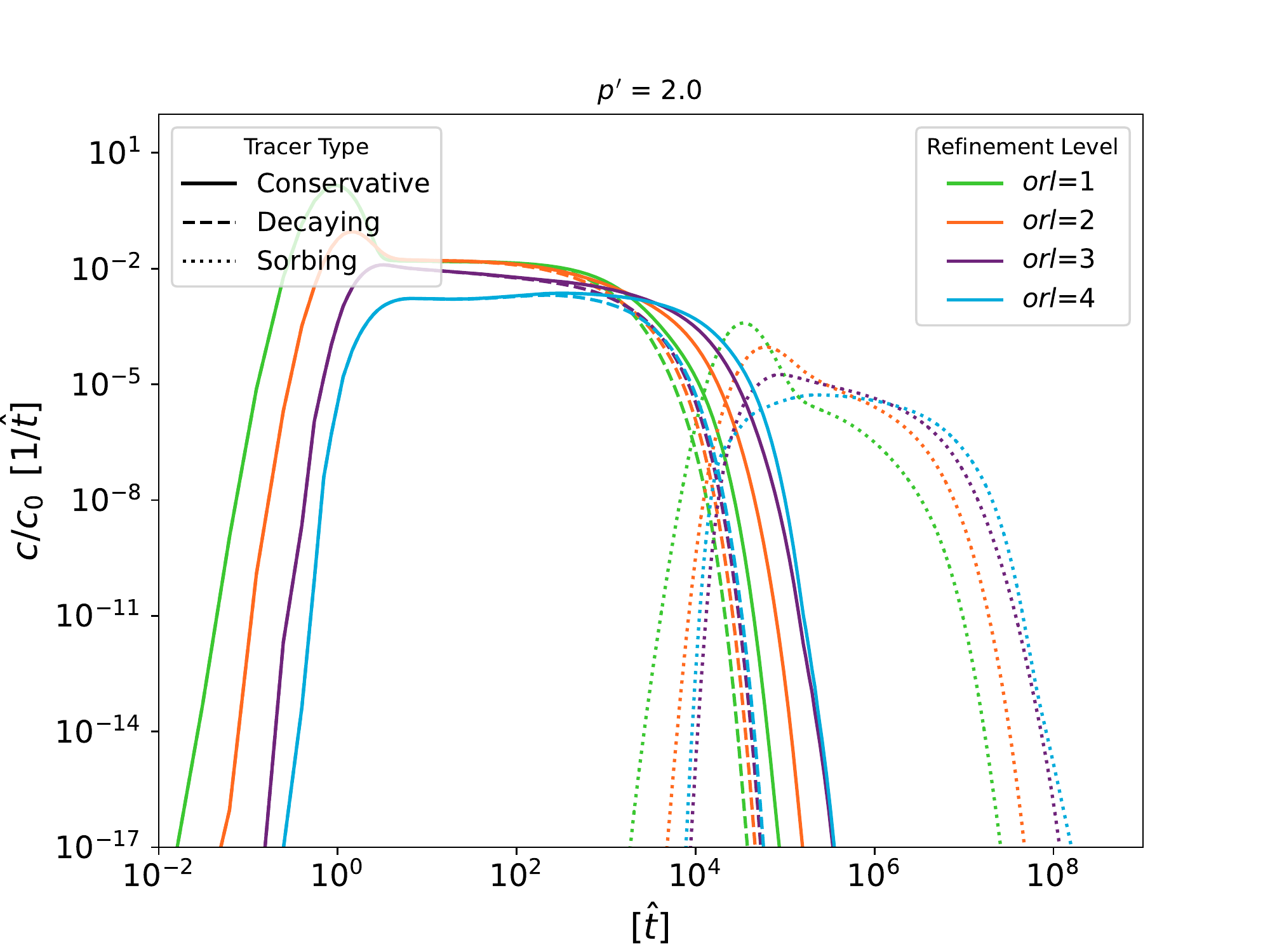}}}
    \subfloat[]{\label{subfig:p2}{\includegraphics[width=0.54\columnwidth]{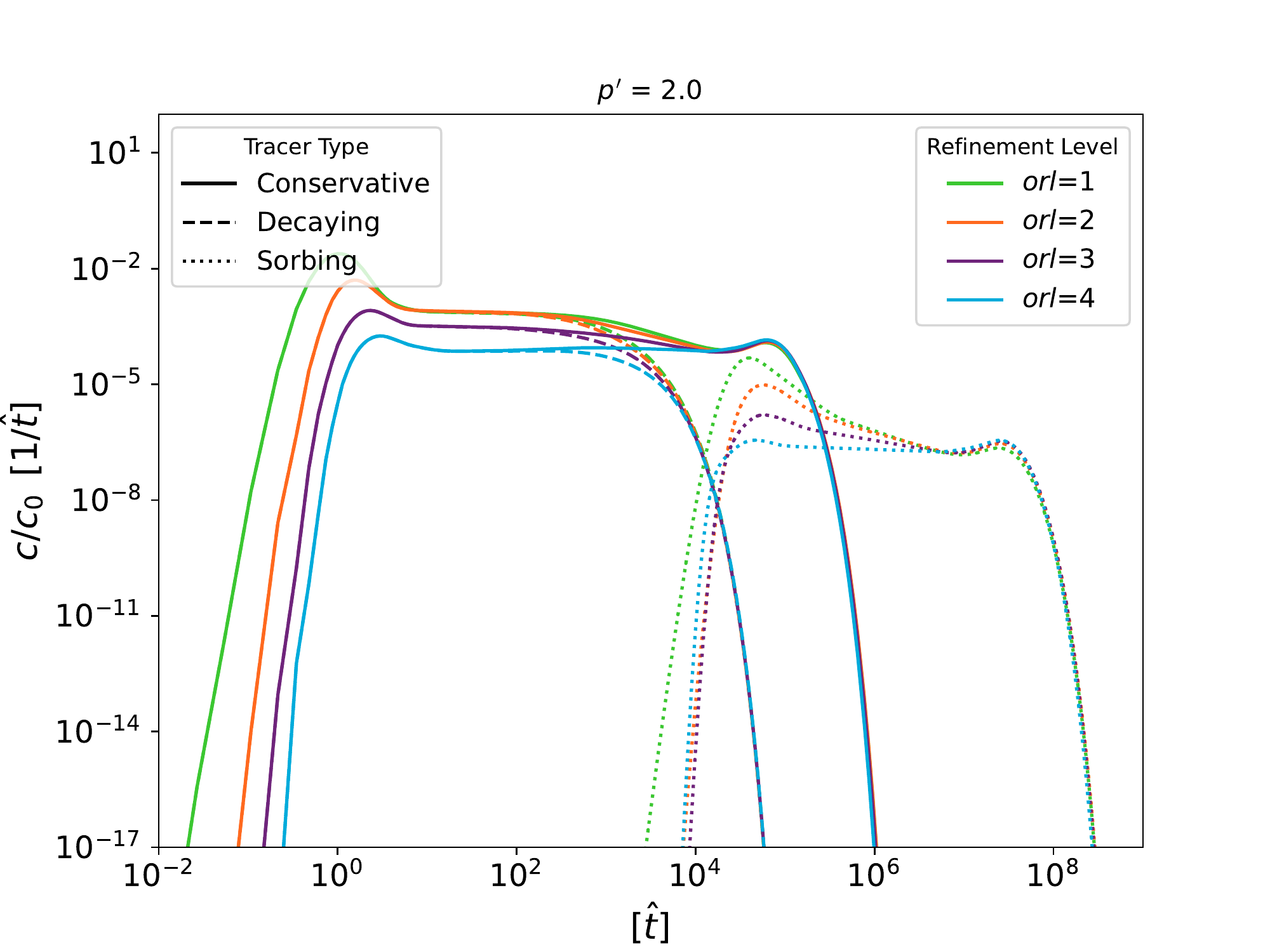}}}\\
    \caption{Breakthrough curves of Eulerian tracer in a fracture network with density $p^\prime = 2.0$. $c$ is the concentration of the tracer at time $\hat{t}$, and $c_0$ is the total amount of tracer in the pulse. We examine the solute transport for fracture networks with (a) isolated fractures retained, and (b) isolated fractures removed.}
    \label{fig:p_2_btc}
\end{figure}

Figure~\ref{fig:p_2_btc} shows the BTCs for $p^\prime = 1.5$.
Here, convergence of the BTC with increasing \orl is much clearer than in the other cases, especially for the case where isolated fractures are removed. 
Although the case where isolated fractures are retained has more cells overall, the local false connections and dispersion influences convergence of the breakthrough curves. 

\section{Matrix Permeability Variations}

\begin{figure}[htb!]
     \centering
     \subfloat[]{\includegraphics[width=0.54\textwidth]{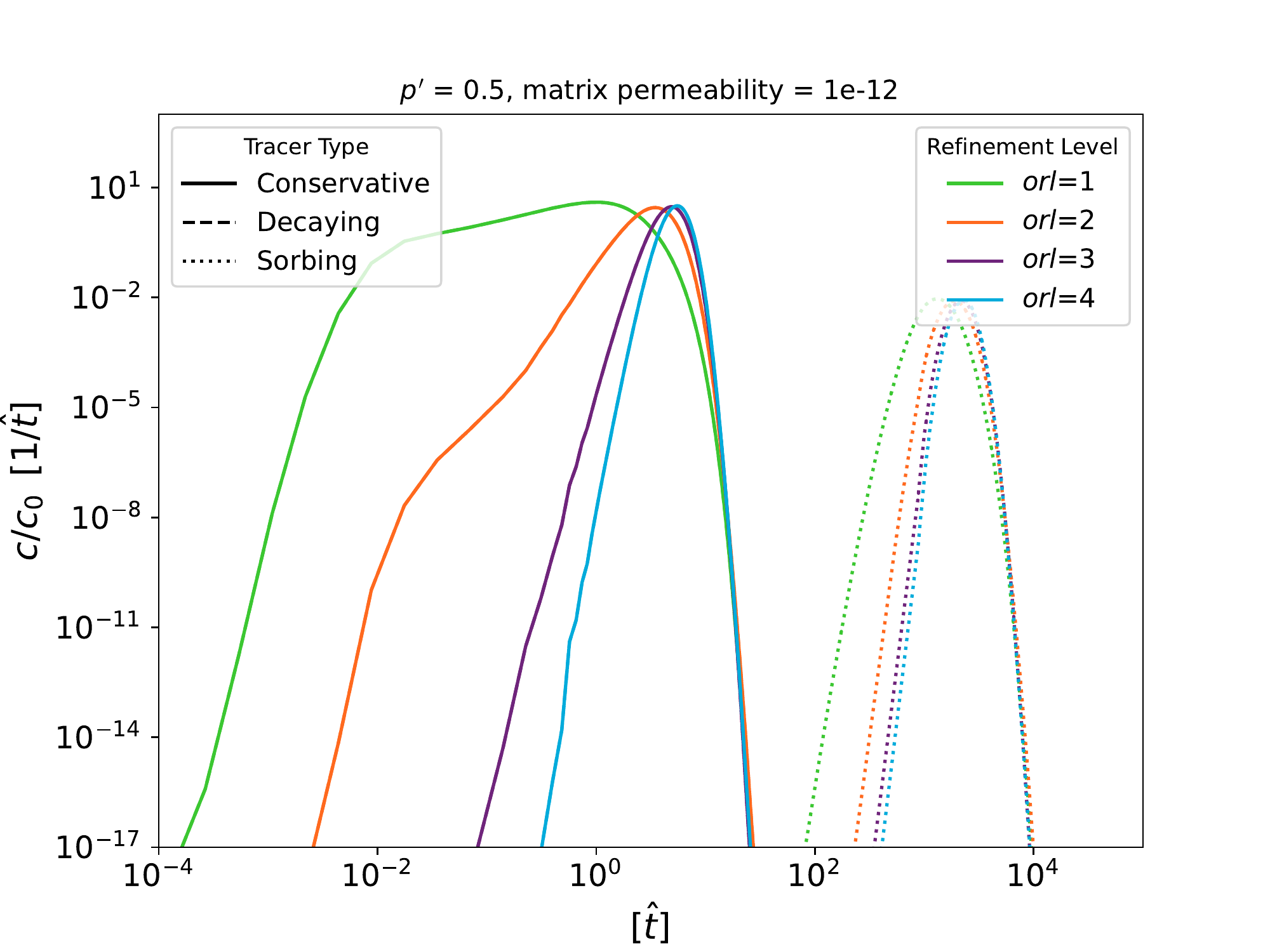}}
     \subfloat[]{\includegraphics[width=0.54\textwidth]{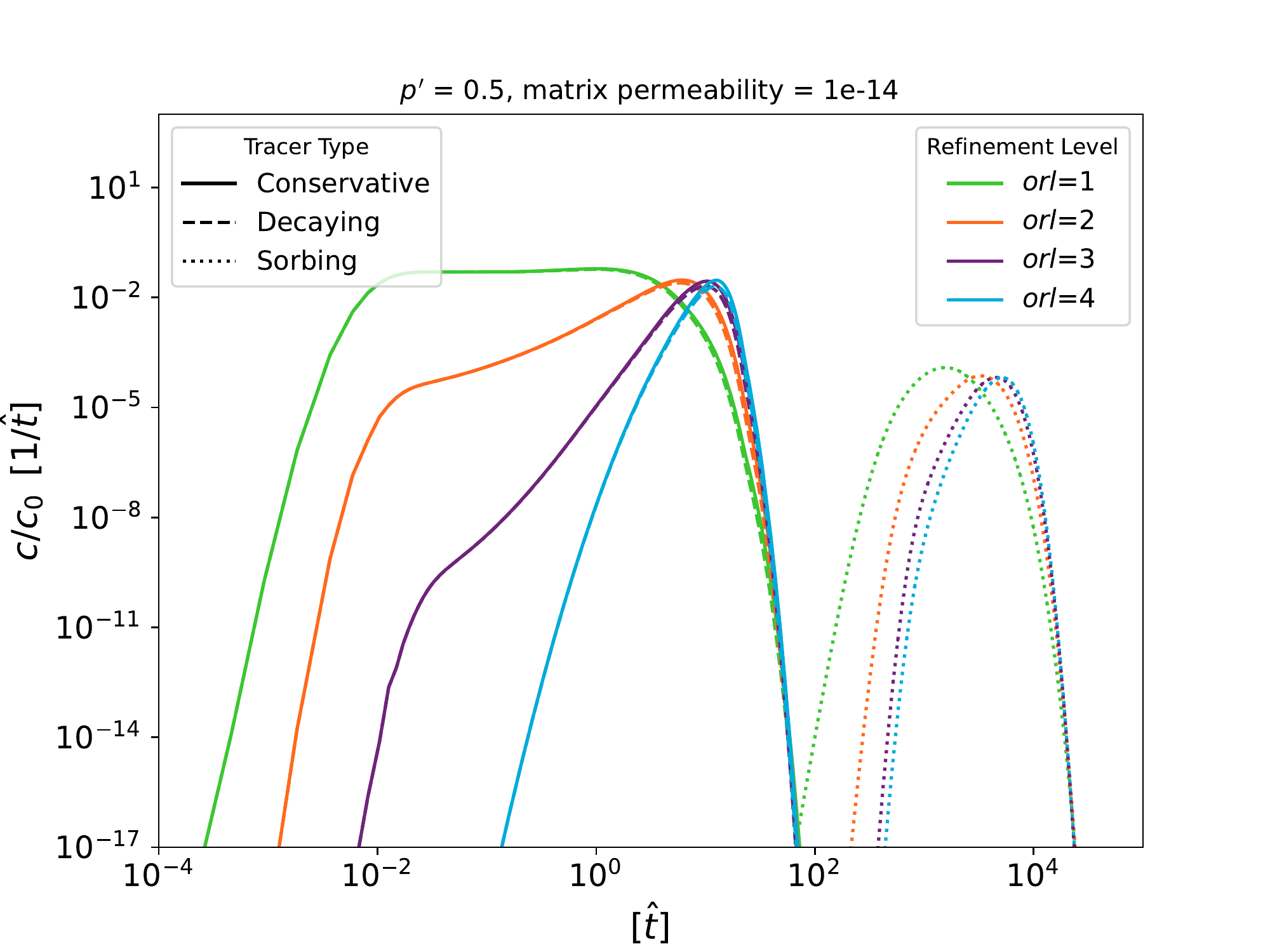}} \\
     \subfloat[]{\includegraphics[width=0.54\textwidth]{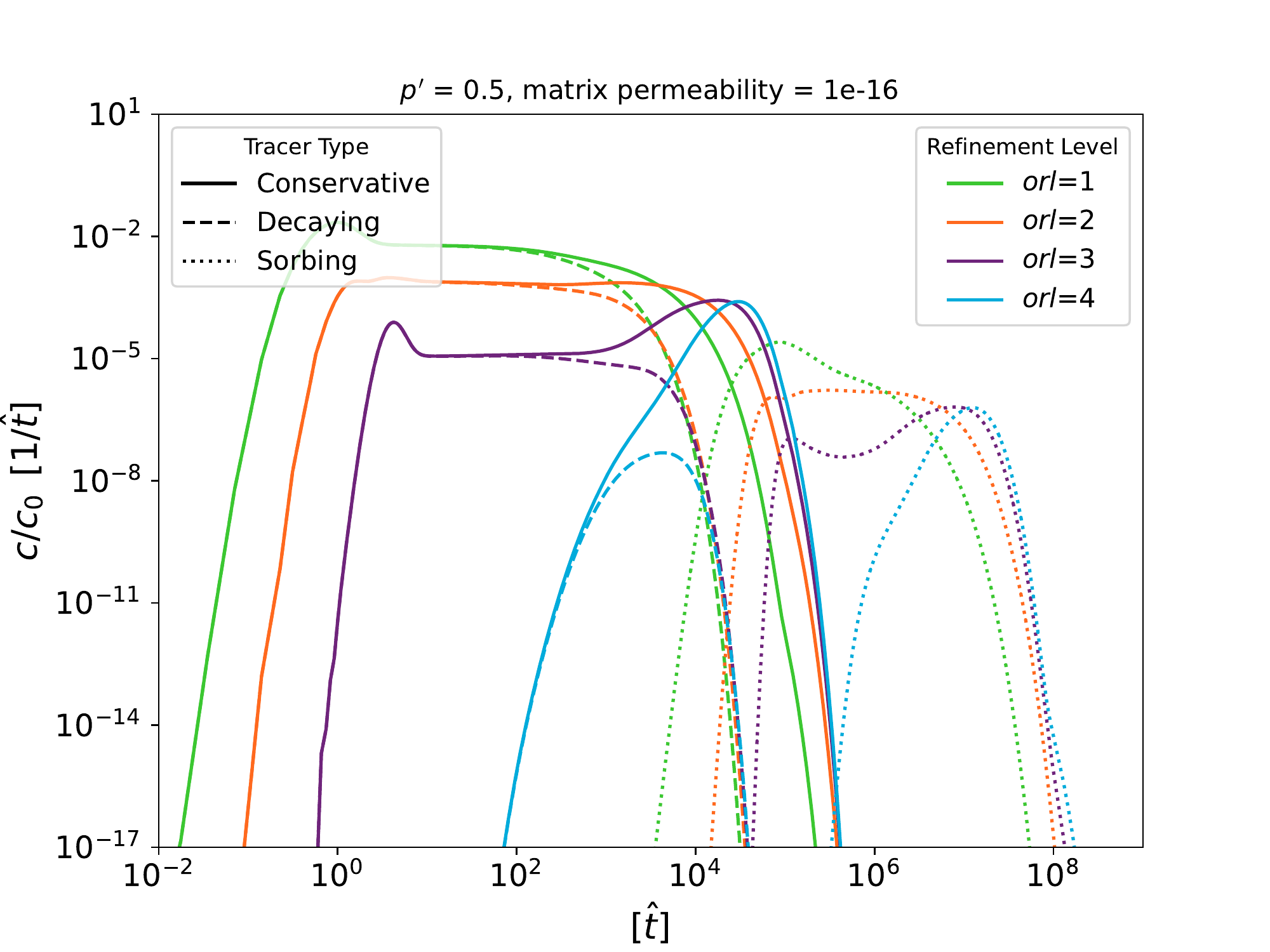}}
     \subfloat[]{\includegraphics[width=0.54\textwidth]{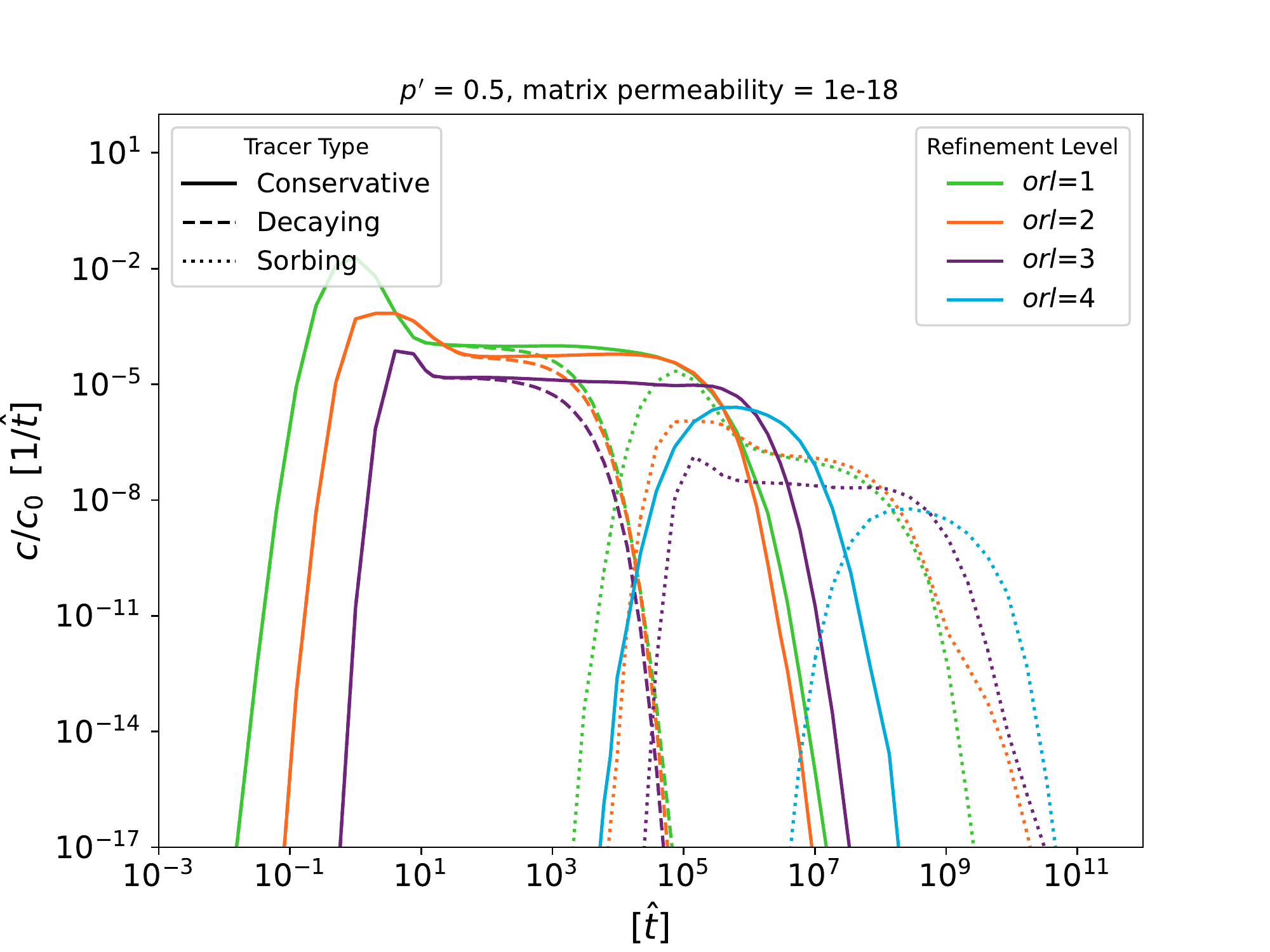}}\\
     \caption{BTCs for matrix permeability values for a fracture network with density $p^\prime = 0.5$. (a)  $k_m = 10^{-12} \text{m}^2$ (b) $k_m = 10^{-14} \text{m}^2$ (c) $k_m = 10^{-16} \text{m}^2$ (d) $k_m = 10^{-18} \text{m}^2$}.
    \label{fig:btc_mat_perm_p05}
\end{figure}

\begin{figure}[htb!]
     \centering
     \subfloat[]{\includegraphics[width=0.54\textwidth]{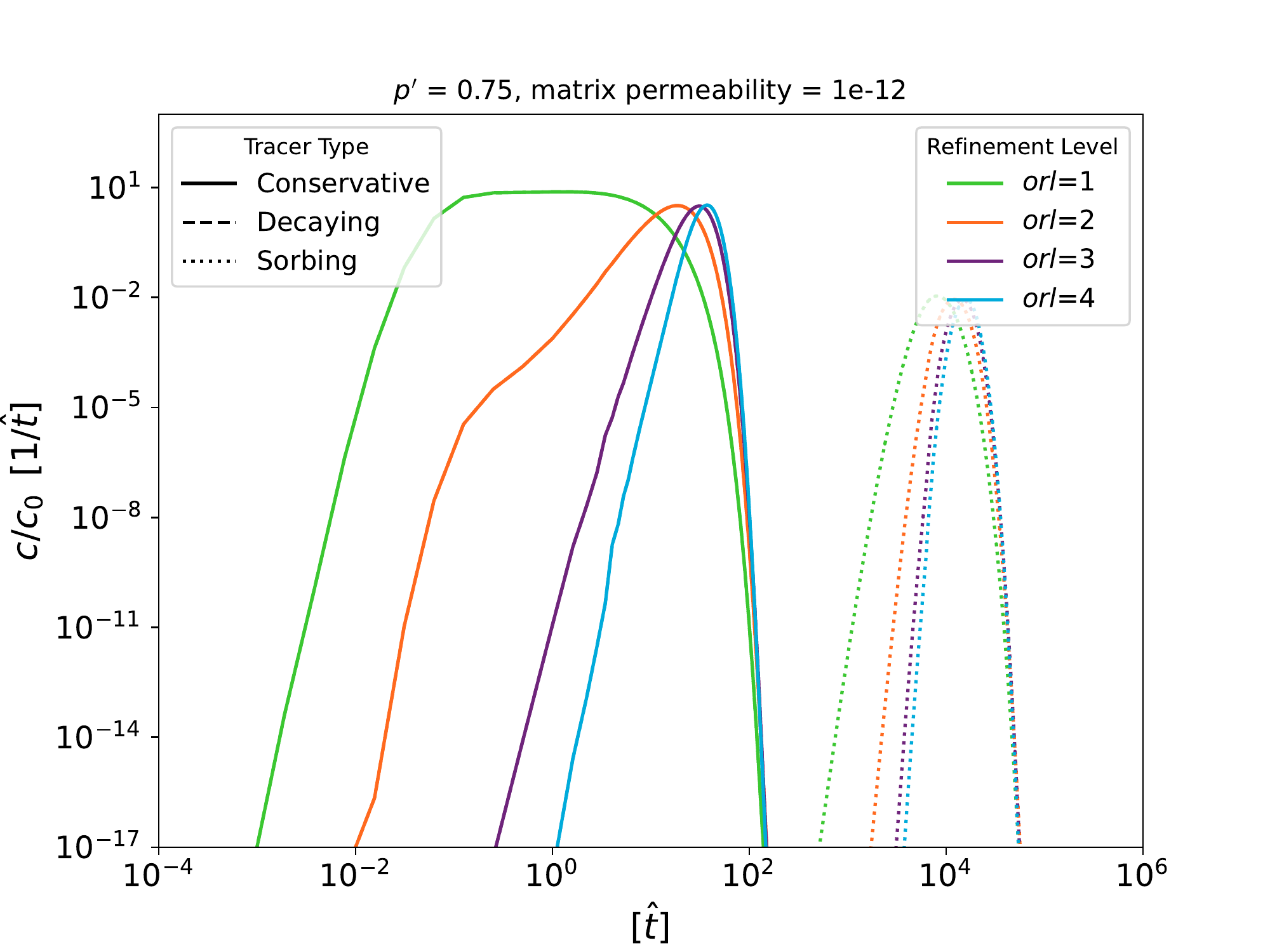}}
     \subfloat[]{\includegraphics[width=0.54\textwidth]{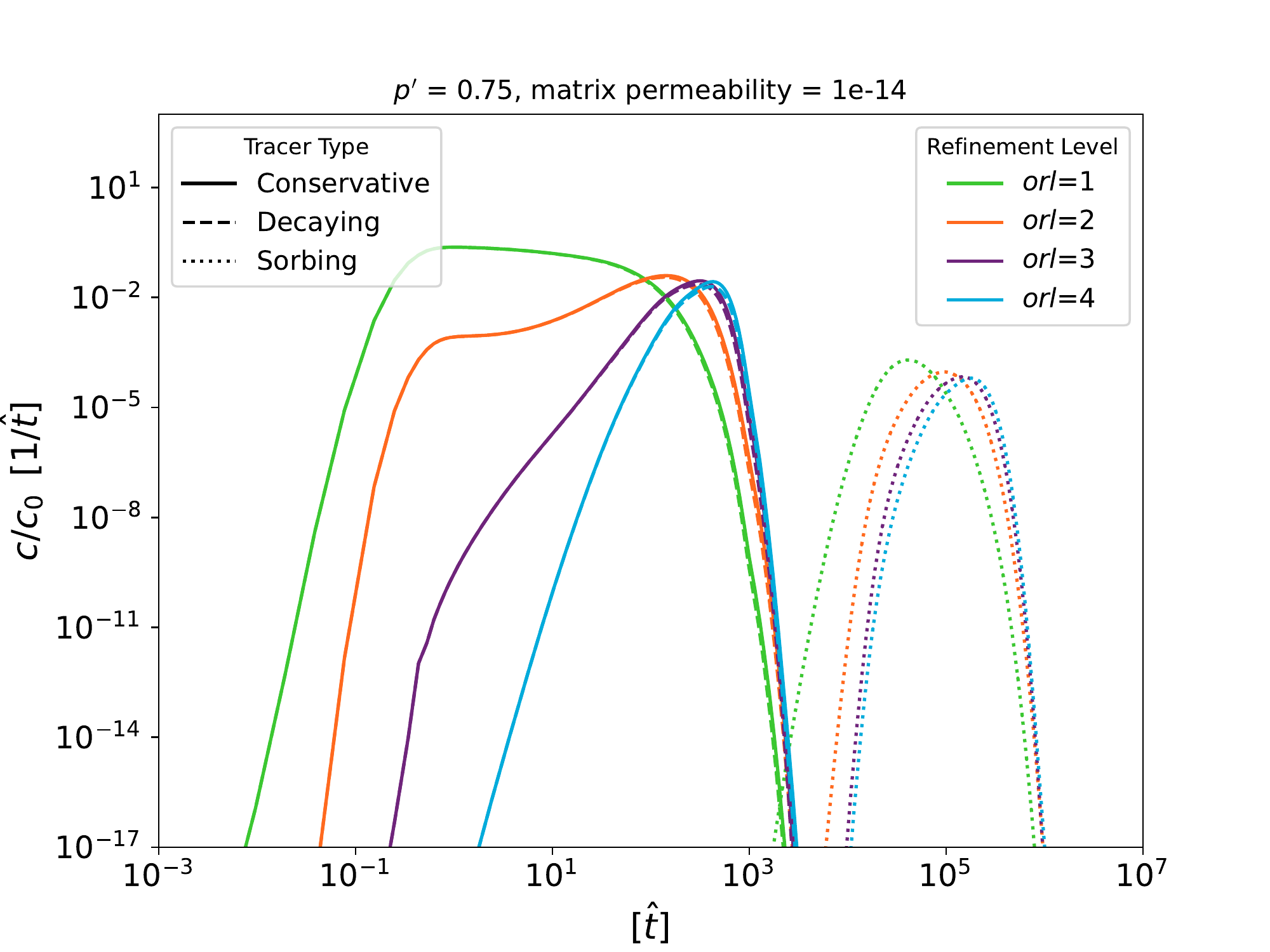}} \\
     \subfloat[]{\includegraphics[width=0.54\textwidth]{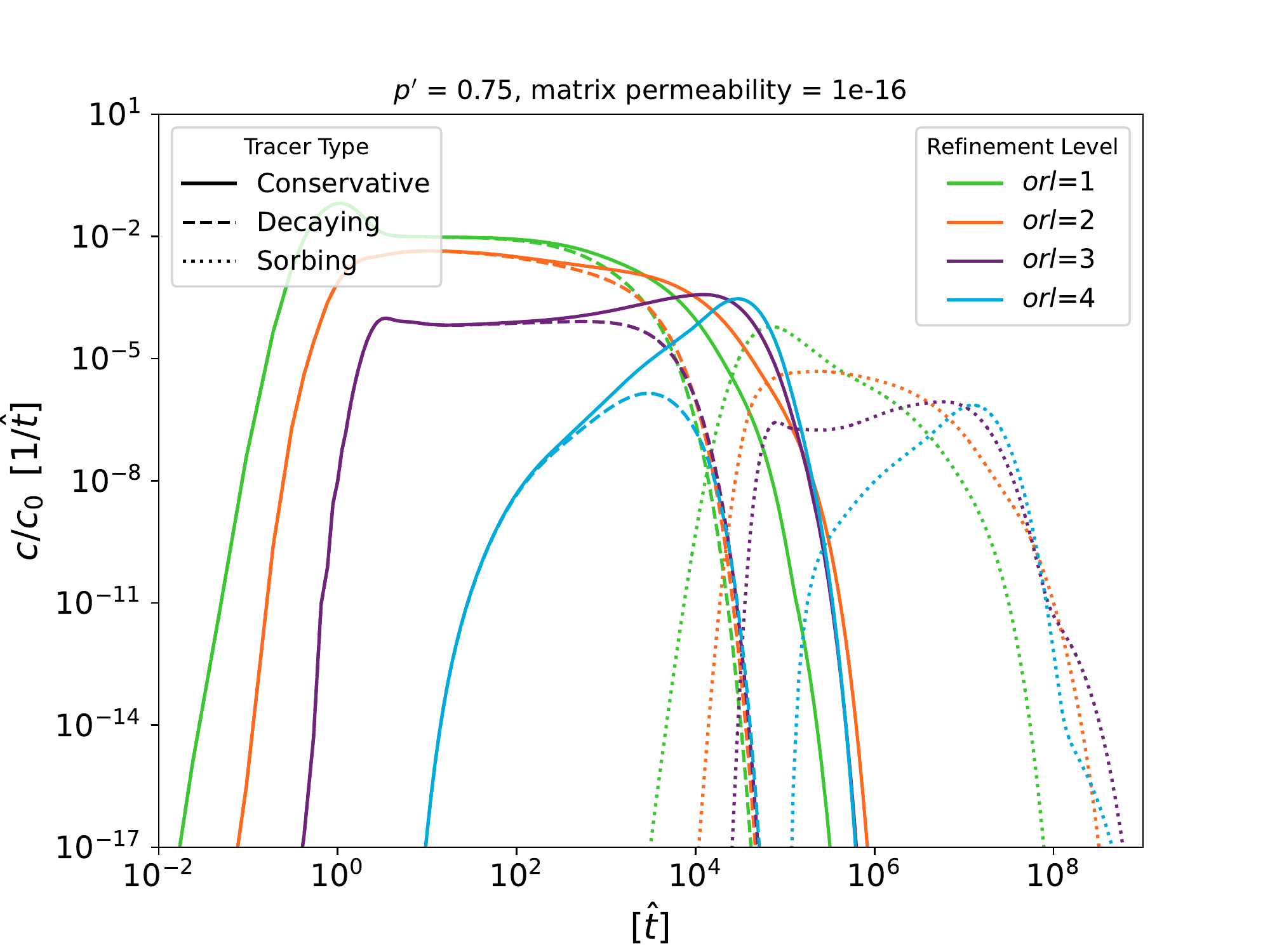}}
     \subfloat[]{\includegraphics[width=0.54\textwidth]{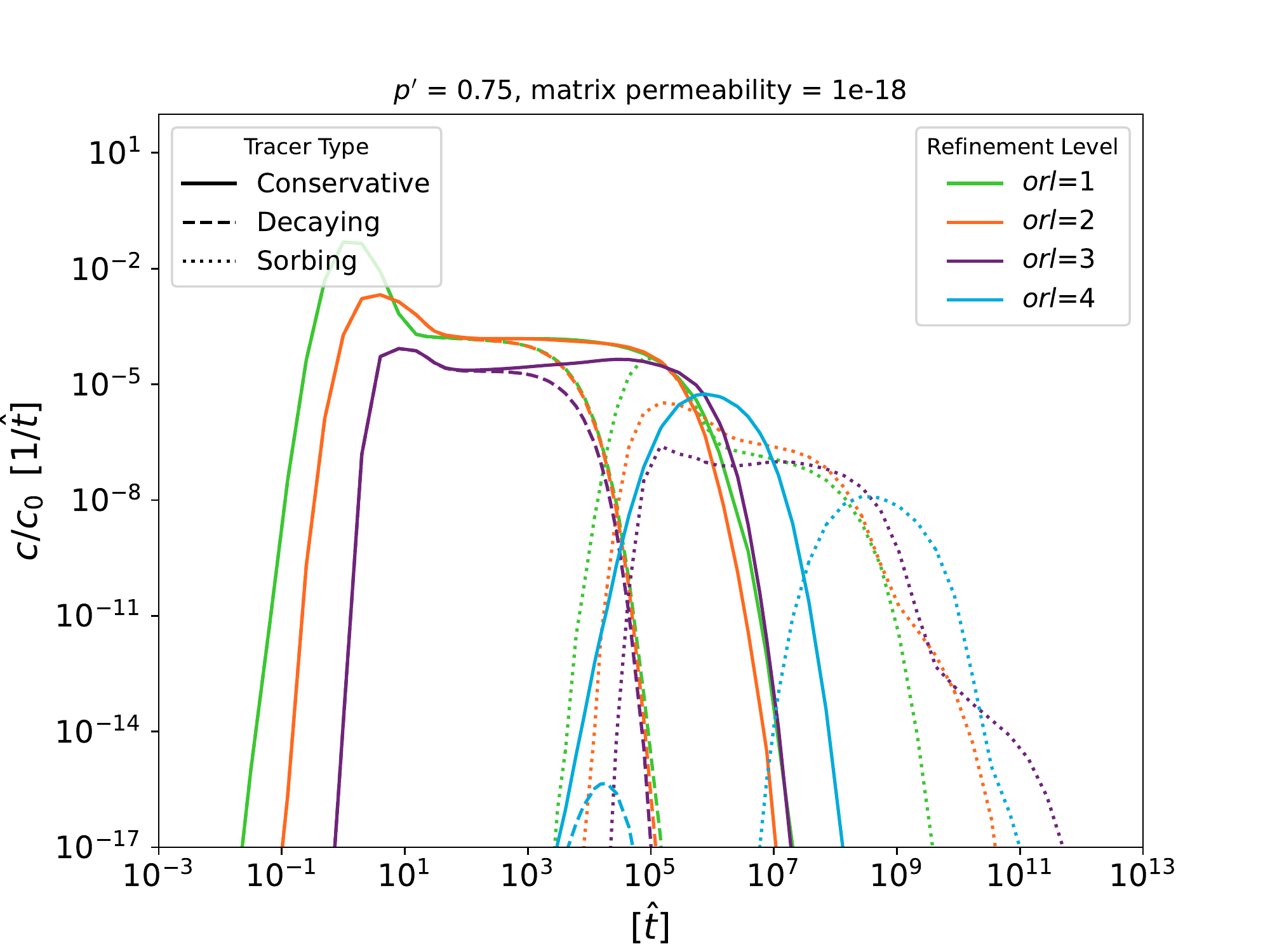}}\\
     \caption{BTCs for matrix permeability values for a fracture network with density $p^\prime = 0.75$. (a)  $k_m = 10^{-12} \text{m}^2$ (b) $k_m = 10^{-14} \text{m}^2$ (c) $k_m = 10^{-16} \text{m}^2$ (d) $k_m = 10^{-18} \text{m}^2$}.
    \label{fig:btc_mat_perm_p075}
\end{figure}

Figure~\ref{fig:btc_mat_perm_p05} and \ref{fig:btc_mat_perm_p075} show the breakthrough curves for \pp = 0.5 and \pp = 0.75, respectively, for the four matrix permeability values that we considered.
Recall that the domain for the $k_m = 10^{-12} \text{m}^2$ values are relative homogeneous, and that is reflected in the BTC. 
The BTC here, appear more like a pulse traveling through a low-variance porous media than and fractured system. 
For \orl = 1, the pulse is wide, and then narrows with rising \textit{orl}.
The permeability is sufficiently high that we do not see the effects the decay rate in the decaying tracer. 
As the matrix permeability rises, the influence of the embedded fractures can be seen in the form of the false early breakthroughs due to the network-scale false percolation. 
These results suggest that not only is the topology of the UDFM mesh important, but there is an interplay between the heterogeneity of the domain and topology.

\section{Summary, Discussion, and Conclusions} \label{sec:conclusion}
We have provided a set of numerical simulations to characterize the influence of mesh refinement level on flow and transport properties in continuum representations of fractured porous media. 
We generated a set of generic three-dimensional fracture networks at various densities, where the radii of the fractures were sampled from a truncated power-law distribution, with parameters loosely based on field site characterizations. 
We considered five network densities, defined using a dimensionless version of density based on percolation theory. 
Two values were below the critical percolation value, one at the value, and two above. 
We considered two sub-cases for each density value.
In the first, we considered the network with all of the generated fractures.
In the second, we only considered non-isolated fractures and clusters relative to the flow conditions. i.e., we retained only clusters of fractures that connected the inflow and outflow boundaries.
Once the networks were generated, we upscaled them into a single continuum model using the UDFM method presented by~\citet{sweeney2020upscaled}.
We considered four octree refinement levels of \orl = 1,2,3, and 4.
We considered steady isothermal pressure-driven flow through each domain and then simulated conservative, decaying, and adsorbing tracers using a pulse injection into the domain. 

Keeping the isolated fractures resulted in more false connections when compared to the cases where isolated fractures were removed, and thus more refinement levels were needed to resolve the true structure of the fracture network.
This was most apparent for the low densities ($p^\prime=0.5$ and $p^\prime=0.75$), where we found that even a refinement level of \orl=3 was not sufficient to eliminate the false connections in the network and the UDFM mesh percolated, when in fact, the DFN did not.
These global false connections dramatically affected the values of effective permeability.
The estimates provided by \orl=1 and 2, were multiple orders of magnitude higher than that provided by \orl=4, which did not percolate. 
In the cases where the topology matched between the UDFM model and DFN, $p^\prime \geq 1$, the differences in the estimates were smaller.
Likewise, when we considered a range of effective permeability values, we saw that not only is the topology of the UDFM mesh important, but there is an interplay between the heterogeneity of the domain and topology. 
For cases with lower matrix permeability, the error in effective permeability estimations was much higher than in the case with higher matrix permeability. 

In terms of transport, the conservative tracer breakthrough curves for \orl=1,2,3 for $p^\prime=0.5$, and $p^\prime=0.75$ showed erroneous peaks caused by artificially connected pathways in the system. 
Once the topology of the UDFM mesh and DFN match, then only later time breakthrough is observed, which is consistent with the physical system. 
When increasing the density of the fracture network, the number of possible percolation paths increased and thus the effect of the false connections to the shape of the breakthrough curves decreased. 
While the global topology matched, there were still changes in local connectivity of the mesh, and more dispersion in the breakthrough curves was observed. 
We observed a dual peak structure, where the first peak was slightly higher than the second one, that is shortly before most of the solute had left the domain.
The first peak was due to the flow and transport through the fracture cells, while the second peak was due to transport through the matrix.
The breakthrough curves for all refinement levels and densities decayed sharply after the second peak and overlapped for all fracture networks above the percolation threshold.
For the decaying tracer, the breakthrough curves are narrower and in most cases, disappeared from the system prior to the expected matrix dominated breakthrough. 
In the case of the non-percolating network, the decaying tracer never broke through the matrix at the expected time due to the decay rate. 
Otherwise, the decaying tracer behaved like the conservative tracer. 
The sorbing tracer showed delayed first arrival times (approx. $10^4$) in comparison to the conservative and decaying tracers for the percolating networks. 
When looking at the non-percolating networks, we observed a drastic increase of the first arrival time depending on the refinement level. 
For example, the first arrival time was at about $10^4$ for refinement level \orl=1, and up to $10^6$ for refinement levels \orl=3 and \orl=4. 

In the case of higher matrix permeability, the BTC appear more like a pulse traveling through a low-variance porous media than a fractured system.
For \orl = 1, the pulse is wide, and then narrows with rising \textit{orl}.
As the matrix permeability rises, the influence of the embedded fractures can be seen in the form of the false early breakthroughs due to the network-scale false percolation. 
These results again emphasize that not only is the topology of the UDFM mesh important, but there is an interplay between the heterogeneity of the domain and topology.

The results of the simulations point to a few key considerations when upscaling fracture networks. 
Foremost, selecting a mesh resolution so that the global topology of the upscaled mesh matches the fracture network is essential. If the upscaled mesh has a connected pathway of fracture (higher permeability) cells but the fracture network does not, then the estimates for effective permeability and solute breakthrough are incorrect. Previous work that studied the consistency of ECPMs relative to DFNs did not consider these global topological changes in their analyses~\citep{jackson2000,kottwitz}. To prevent such situations, identifying percolation in the fracture network and upscaled mesh is an essential part of the modeling process. Such queries can be easily performed using graph theory software such as networkX~\citep{hagberg2008}, among others.
Local false connections also impact transport behavior, but to a smaller degree than if the global connectivity is incorrect, which is consistent with previous work~\citep{jackson2000,kottwitz}. These small short circuits in the system, which occur more frequently if isolated fractures are retained,  increase dispersion of the solute plume. We did not discriminate between dispersion caused by false connections themselves and numerical dispersion caused by mesh size and numerical scheme, but this would be a problem worth investigating in future work.
False connections cannot be eliminated entirely, but they can be managed with mesh resolution. 
While using a uniform mesh resolution is simple, adopting that methodology can lead to intractable numbers of small cells. 
On the other hand, adopting octree meshing to obtain sufficient levels of refinement leads to fewer computational cells (up to a 90\% reduction in overall cell count), but it is more laborious to create the mesh. 

The computational representation of fractured porous media remains a complex part of subsurface flow and transport modeling. 
While numerical models are rapidly advancing with the introduction of multi-dimensional DFM models, kinematic DFN, embedded fracture matrix models, and coupled thermo-hydro-mechanical-chemical models~\citep{ahmed2015control,ahmed2015three,antonietti2016mimetic,arraras2019mixed,boon2018robust,berrone2018advanced,davy2010likely,davy2013model,Flemisch2016,fum,hyman2022flow,ODS,lagrange,lavoine2020discrete,manzoor2018interior,sandve2012efficient,Schwenck2015,thomas2020permeability,viswanathan2022from}, mesh generation remains a key step for all modeling endeavors. 
Therefore, continued characterization of how meshing techniques impact physical phenomena beyond the standard convergence of the numerical scheme remains requisite for modeling endeavors.
Extensions of this research should focus on how the influence of meshing methods interplay with various upscaling techniques as well as the aforementioned modeling methodologies.

\begin{acknowledgements}
JDH, MRS, and HSV thank the Department of Energy (DOE) Basic Energy Sciences program (LANLE3W1) for support.
JDH, MRS, and HSV also gratefully acknowledges support from the LANL LDRD program office Grant Number \#20220019DR.
AP acknowledges the Center for Non-Linear Studies at LANL.
Los Alamos National Laboratory is operated by Triad National Security, LLC, for the National Nuclear Security Administration of U.S. Department of Energy (Contract No. 89233218CNA000001).
This work has been partially funded by the Spent Fuel and Waste Science and Technology (SFWST) Campaign of the U.S. Department of Energy Office of Nuclear Energy.
Sandia National Laboratories is a multi-mission laboratory managed and operated by National Technology and Engineering Solutions of Sandia, LLC., a wholly owned subsidiary of Honeywell International, Inc., for the U.S. Department of Energy’s National Nuclear Security Administration under contract DE-NA-0003525.
The views expressed in the article do not necessarily represent the views of the U.S. Department of Energy or the United States Government.
Assigned LA-UR-23-20307.\\
\end{acknowledgements}


\section{Data Availability}
The datasets generated during and/or analysed during the current study are available from the corresponding author on reasonable request.

\section{Declarations}
The authors have no competing interests to declare that are relevant to the content of this article.




\bibliographystyle{spbasic}      
\bibliography{refs}

\end{document}